\documentclass[english,11.5pt]{article}
\usepackage{verbatim}
\usepackage{times}
\usepackage{helvet}
\usepackage{courier}
\usepackage{eurosym}
\usepackage{graphicx}
\usepackage{amsmath,bm}
\usepackage{amsthm}
\usepackage{amsfonts}
\usepackage{tikz,amssymb}
\usepackage{xr-hyper}
\usepackage{authblk}
\usepackage{bm}
\usepackage{comment}
\usepackage{appendix}
\usepackage{graphicx}
\usepackage{endnotes}
\usepackage{tikz}
\usepackage{booktabs}
\usepackage{subcaption}
\usepackage{url}
\usepackage{hyperref}
\usepackage[numbers]{natbib}
\usepackage
[
a4paper,
left=1in,
right=1in,
top=1in,
bottom=1in,
]{geometry}
\providecommand{\keywords}[1]{\textbf{\textit{Keywords ---}} #1}
\usepackage{setspace}
\newcommand{\IGNORE}[1]{}

\let\footnote=\endnote

\usepackage{longtable}

\begin{document}

\doublespacing

\title{\color{black} \vspace{-5.5ex}Inferring Behind-the-Meter Demand Components from Net Load via Inverse Optimization}
\date{\vspace{-6ex}}

\author[1]{Adri\'an Esteban-P\'erez}
\author[2]{Derek Bunn}
\author[3]{Yashar Ghiassi-Farrokfal}

\affil[1]{Department of Computer and Systems Sciences \& Digital Futures,
Stockholm University,
P.O. Box 1073, SE-164 25 Kista, Sweden\\
\texttt{adrian.esteban@dsv.su.se}}

\affil[2]{London Business School, London NW1 4SA, United Kingdom\\
\texttt{dbunn@london.edu}}

\affil[3]{Department of Technology and Operations Management,
Rotterdam School of Management, Erasmus University,
Rotterdam 3062 PA, The Netherlands\\
\texttt{y.ghiassi@rsm.nl}}
\date{\vspace{-6ex}}

\maketitle

\maketitle

\begin{abstract}
Understanding and forecasting electricity demand responses to price signals is essential to support the operational decisions of system operators, retailers, and regulators. Traditional time series and machine learning approaches have performed well under stable consumption patterns. However, the rise of active consumers with flexible assets, such as solar-plus-storage systems and electric vehicles, introduces new complexities. These consumers exhibit behind-the-meter behavior that is generally unobservable to system operators and retailers, who can typically only measure net demand at the boundary meters which are located at the connection points to the grid. Net demand reflects the aggregate of baseload consumption, flexible load shifting, and self-generation, all of which may respond nonlinearly to market prices. As these unobservable components grow, existing forecasting models may fail to capture the true dynamics of consumer behavior. This work develops and implements a data-driven inverse optimization framework to reveal behind-the-meter consumption patterns using only net demand data. By analyzing net demand as revealed at the grid connection point, we estimate parameters for a latent optimization model, enabling predictions that infer the unobservable components. This enables more accurate point and probabilistic forecasting of electricity demand under varying price signals. Thus, it provides actionable insights for system operators to improve local congestion management and flexibility extraction, as well as for retailers and aggregators, who could thereby optimize operations and tariffs to leverage consumer flexibility. We validate the methodology using real-world data, demonstrating improved forecasting accuracy over leading machine learning and time-series benchmarks.
\end{abstract}
%%%%%%%%%%%%%%%%%%%%%%%%%%%%%%%%%%%%%%%%%%%%%%%%%%%%%%%%%%%%%%%%%%%%%%

% Text of your paper here

\section{Introduction}\label{sec:Intro}

Understanding and predicting the electricity demand responses to prices are critical for decision making of system operators (SO), retailers and regulators. For system operators, accurate demand forecasts are essential to ensure that energy supply meets demand within the grid's capacity limits and other physical constraints \citep{shahidehpour2002market}. For electricity retailers and aggregators, these predictions are vital for revenue management, cost control, and tariff design to provide better pricing for consumers \citep{karthikeyan2013review}, whilst for regulators and policy-makers, promoting efficient consumer engagement has become an important element in their net zero ambitions \citep{pinson2014benefits}. Electricity demand forecasting at the consumer level has been widely researched and, in practice, had until recently reached acceptable levels of accuracy, largely due to the relatively stable nature of consumer behavior \citep{hatton2015statistical}. However, recent developments associated with the energy transition have radically altered all elements of the electricity supply and value chain.  
{\color{black}Furthermore, the global energy crisis, beginning during 2021-22, which followed from both the problems of economic recovery from the COVID-19 pandemic and the Russian war, focused attention on energy independence and consequent the role of increased energy efficiency  \citep{gajdzik2024influence,farghali2023strategies}}

These changes have been driven by substantial subsidies for decarbonization and the emergence of end-user generation. As a result, system operators have recognized an emergent need for energy {\em flexibility} to help balance supply and demand \citep{mathieu2024new,pinson2014benefits}.
{\color{black} Thus, it has been argued that energy saving is no longer merely an environmental choice but a necessary adaptation during periods of scarcity, where the ability to shift or reduce load may be critical to maintaining grid stability \citep{farghali2023strategies}.} Flexibility in this context mainly refers to the ability to adapt energy consumption patterns, such as shedding demand during periods of high energy prices or grid stress, or shifting demand to times when energy is cheaper or the grid is less stressed. To promote consumer flexibility, retailers have introduced new incentive mechanisms, including Time-of-Use (TOU) pricing and dynamic pricing models, which align flexibility with grid status and/or temporal energy prices \citep{vardakas2014survey}. In response, consumers have increasingly embraced flexibility e.g., by adopting solar-plus-storage systems and opportunistically scheduling their electric vehicle charging, thereby enhancing their ability to manage energy costs and contribute to grid stability \citep{danti2017effects}.

As a result of these changes, accurate estimation of the consumer demand response in the evolving retail electricity markets has become both more critical and more complex. This importance has increased because energy procurement is now more challenging due to the added volume and uncertainty, pushing the traditional grid closer to its limits and raising the risk of congestion \citep{koliou2015quantifying}. The complexity has grown because, unlike in the past where demand was generally not price responsive on an hourly basis, today’s consumer demand, as observed by SOs and retailers at the metering point with the grid, is confounded by new unobservable components.  The essential baseload demand (unresponsive to price) is supplemented with price-responsive flexible demand (including batteries and electric vehicles) and reduced by any end-user generation (e.g., solar). Accurately estimating these components would require retailers and SOs to know the types and capacities of flexible devices owned by consumers and to have access to real-time device-level data on how consumers operate these assets to manage costs. However, this information, known as behind-the-meter (BTM) behavior, is typically hidden from SOs and retailers, as the meters only reveal "net demand".  In fact, not only are they often unaware of the devices owned by consumers—sometimes even including solar panels—but they also may be unable to access or utilize real-time device-level measurements. This data is typically classified as personal information and protected by privacy legislation, for example by the General Data Protection Regulation (GDPR) in the EU \citep{EU2016}, and sometimes by considerations of cybersecurity in smart systems \citep{liu2012cyber}. The difficulty and importance of accurately characterizing BTM  flexibility behavior has motivated increasing attention by industry and researchers on the estimation, optimization, and control of BTM behavior 
\citep{srivastava2024distribution}. 

Consequently, SOs and retailers need to estimate the unobservable behind-the-meter information in the absence of direct measurements. But existing methodologies, such as time-series analysis \citep{choi2020have,dong2017electricity} and machine learning algorithms have generally been applicable only to the observable demand, as metered at the grid connection point (``net demand'') rather than the specific latent components of consumer demand which may be flexing \citep{nghiem2017data,wen2020modified}. To address this methodology gap, we question if the emerging technique of inverse optimization (IO) could offer a promising solution for estimating the unobserved information on these components, thereby potentially enhancing the prediction of observable net demand. IO operates by modeling a consumer's behind-the-meter price response as a latent optimization problem under different pricing schemes and then integrating this into an optimal prediction problem. Intuitively, this method reverses the typical optimization methodology: rather than determining optimal actions according to a constrained objective function, it seeks to infer the parameters of the model on the basis of revealed actions, assuming a unobservable latent optimization approach by the consumers. This approach therefore has the potential to capture nonlinear consumer responses by distinguishing between their flexible, inflexible, and self-generation components of demand, to the extent that they are specified in a latent model of  optimal demand response. Whether this decomposition then leads to improved net demand forecasts is the open research question that we address.  

As such, our work introduces a data-driven Inverse-Optimization method that enables SOs and retailers to infer demand components and behind-the-meter flexibility of consumers, helping inform decisions around integrating consumer flexibility into grid operations (for SOs) and {\color{black}{benefit}} maximization strategies (for retailers). Crucially, our approach preserves data privacy and avoids the need for real-time, device-level measurements. 
 In this paper, we contribute to this effort by helping retailers and SOs estimate unobservable information in demand behavior.   The key contributions of this paper are therefore summarized as follows:

\begin{itemize}
    \item {\bf Domain Contributions:} This paper makes a substantial contribution to characterizing the unobserved components of electricity demand response, being, as far as we are aware, the first to do so in the absence of device-level, real-time measurements. We accomplish this by developing an application of inverse optimization (IO) to this critical issue, a method previously used in other fields but not yet in retail electricity. Unlike conventional approaches that depend on direct observations of device-level data, this new method leverages net demand data to estimate unobservable flexibility parameters within a latent optimization model. 

    \item {\bf Experimental Results:} To validate the performance of the proposed methodology, we use real-world data from Kaggle, which provides access to detailed, device-level, behavior of a US household with the aim of supporting multiple replication studies. Applying inverse optimization (IO) to this data, we find that the estimated flexible and inflexible components of demand align well with actual device behavior, confirming the plausibility of our approach. 

    \item{\bf Practical Implications:} This methodology allows retailers and system operators to gain a deeper understanding of behind-the-meter flexibility without compromising data privacy or requiring real-time device-level measurements. It can help them forecast better and enhance their incentive mechanisms, resulting in monetary benefits for themselves and consumers through more efficient green energy management.

\end{itemize}

The rest of the paper is organized as follows.  Section~\ref{sec:background} reviews the related literature. In Section~\ref{sec:model_intro}, we introduce the problem. In Section~\ref{sec:IO}, we provide the proposed IO model and its tractable reformulation. Section~\ref{sec:case_study} reports the empirical results and Section~\ref{conclusions} concludes the paper.

\section{Background research}\label{sec:background}
Our work relates to three main research streams: electricity demand forecasting models, elastic demand analysis and behind-the-meter estimation, and inverse optimization methodologies. We discuss each of these streams separately.

\subsection{Electricity demand forecasting models}
Electricity demand forecasting at the consumer level has been widely researched and, in practice, had until recently reached acceptable levels of accuracy, largely due to the relatively stable nature of consumer behavior \citep{hatton2015statistical}. However, recent developments associated with the energy transition have radically altered all elements of the electricity supply and value chain. 
{\color{black}Furthermore, the global energy crisis, beginning during 2021-22, which followed from both the problems of economic recovery from the COVID-19 pandemic and the Russian war, focused attention on energy independence and consequent the role of increased energy efficiency  \citep{gajdzik2024influence,farghali2023strategies}}

Existing methodologies, such as time-series analysis \citep{choi2020have,dong2017electricity} and machine learning algorithms have generally been applicable only to the observable demand, as metered at the grid connection point (``net demand'') rather than the specific latent components of consumer demand which may be flexing \citep{nghiem2017data,wen2020modified}. A fundamental issue with the regression methods, both statistical and machine learning, is that the training (estimation) relies on regressors rather than being driven by actual end-user's price responses. This has been emphasized by  \cite{chen2024flexenvelope,hekmat2023data}. 

{\color{black}In this context, \cite{ghelasi2025data} argue that purely data-driven models often lack the physical and economic structure necessary to faithfully represent price formation. By proposing a data-driven merit order model, they demonstrate that embedding fundamental principles—akin to the optimization logic underlying IO—allows the model to learn structural parameters directly from market observations. This hybrid approach highlights that the explicit representation of an agent's decision function is crucial to overcoming the limitations of rigid elasticity assumptions in many conventional methods and the lack of interpretability in most machine learning \citep{ruan2021estimating, ponocko2018forecasting}.}

\subsection{Elastic demand analysis and behind-the-meter estimation}
The complexity of demand estimation has grown because today’s consumer demand is confounded by new unobservable components: essential baseload demand is supplemented with price-responsive flexible demand and reduced by end-user generation. System operators have recognized an emergent need for energy {\em flexibility} to help balance supply and demand \citep{mathieu2024new,pinson2014benefits}. {\color{black} Thus, it has been argued that energy saving is no longer merely an environmental choice but a necessary adaptation during periods of scarcity, where the ability to shift or reduce load may be critical to maintaining grid stability \citep{farghali2023strategies}.}

Various attempts to decompose ``behind-the-meter'' consumer demand into its inflexible baseload and flexible, price responsive, components have ranged from statistical and optimization techniques to machine learning approaches \citep{dobakhshari2018contract,hatton2015statistical,lei2020baseline,sun2019clustering,wen2020modified,zhang2015cluster}. Regarding the flexible demand component, whilst its price response is confounded with other behavioral and weather-dependent factors, probabilistic methods including Bayesian inference and machine learning have provided useful insights into these uncertainties \citep{lin2021privacy,valles2018probabilistic,mahdavi2022probabilistic,zhang2022multi}. 

Nevertheless, all these methods encounter significant challenges, particularly in capturing temporal dependencies and handling nonlinear characteristics. Various empirical studies have identified limitations associated with nonlinear effects, delayed responses and a general tendency for over-estimated price responses \citep{an2015transfer,an2016dynamic,cappers2010demand,kirschen2000factoring}. The methods prevalent in existing research struggle with these and other limitations, primarily due to rigid assumptions in the elasticity models \citep{ruan2021estimating}, the lack of interpretability in machine learning \citep{ponocko2018forecasting}, and the misalignment between demand baseline estimation and flexibility components \citep{valles2018probabilistic}. For example, \cite{ruan2021estimating} introduce Siamese LSTM networks to estimate time-varying elasticities through a two-stage process. Similarly, \cite{ponocko2018forecasting} propose a two-step approach that first decomposes load using device-level data, followed by load forecasting with an artificial neural network. In another study, \cite{valles2018probabilistic} utilize a quantile regression method to capture demand flexibility but it requires prior knowledge of the demand baseline and the upward/downward consumer's response. Furthermore, the two-step process can introduce biases in both net load forecasting and the estimation of baseline demand and flexibility estimates.

\subsection{Applications of inverse optimization in power systems}
Inverse Optimization (IO) takes the results of a presumed optimization process to infer the parameters in the formulation of the decision-maker's optimal actions. \cite{chan2023inverse} provide a comprehensive review of its theory and applications. In practice, real data does not usually permit an exact derivation due to noise, measurement errors, or bounded rationality. Thus, IO methodology seeks to minimize fitting errors between observed data and prescribed model results \citep{aswani2018inverse,mohajerin2018inverse}.

Empirical evidence has shown that where a latent agent optimization model is plausible, the IO approach has greater predictive accuracy than machine learning and time-series models with limited data \citep{bian2023predicting,fernandez2021inverse,saez2017short}. Power systems have been particularly fertile for these applications \citep{saez2017short}. {\color{black}This methodological superiority is reinforced by the work of \cite{ghelasi2025data}, who utilize a fundamental optimization structure as a structural regularizer for the learning process, allowing the model to be both predictively powerful and economically coherent.}

Related to the demand response context, \cite{bian2022,bian2023predicting} develop an IO approach using prior model knowledge and gradient descent. \cite{shi2023demand} consider a deep learning method embedded into a bi-level model. The research by \cite{fernandez2021forecasting,fernandez2021inverse, saez2016,saez2017short} proposes a data-driven IO model for forecasting by reducing bi-level optimization to single-level. Then, by relaxing complementary slackness conditions, they solved the problem by employing a heuristic method. \cite{kovacs2021inverse} apply IO to extract parameters via quadratically constrained quadratic programming but expressed concerns about convergence. {\color{black}\cite{vatandoust2023integrated} consider a data-driven IO approach to infer flexibility curves through aggregate forecasts, outperforming XGBoost in the Belgian imbalance market.}

The IO approach developed in this paper integrates forecasting and demand decomposition endogenously within the overall model, rather than as a sequence of separate tasks. This captures complex interactions robustly while maintaining interpretability and uncertainty quantification, offering benefits over conventional separate modeling \citep[Table II]{srivastava2024distribution}. We extend this body of work by developing a computationally efficient reformulation of the IO program to solve a conic mixed integer program. We avoid the need to apply a gradient descent method \citep{bian2022,bian2023predicting}, being aware that gradient-based algorithms have problems handling bi-level formulations with binary variables \citep{zhoulearning2024}. We also avoid relaxing complementary slackness conditions to apply heuristics, as used in previous works \citep{fernandez2021forecasting,fernandez2021inverse, saez2016,saez2017short}. Our method is also able to provide shiftability and sheddability estimates going beyond other approaches \citep{chen2024flexenvelope, hekmat2023data}.

\section{Model}\label{sec:model_intro}

In this section, we present the model. Throughout the paper, bold lower-case letters denote vectors, while standard lower-case letters are
reserved for scalars. We use the  shorthand $[T]=\{1,\ldots, T\}$
 to represent the set of all integers up to $T$. For reference, we also include all notations in Table~\ref{tab-notation} at Appendix~\ref{notation_appendix}.

The model considers the relationship between two parties: the consumer and the service provider. The service provider may be a retailer, aggregator, or network system operator,  the distinction not being important for the general formulation.  Therefore, for simplicity, we collectively refer to them as the system operator (SO). The SO is responsible for managing the consumer's transactions on the basis of net demand information, taken from the consumer's meter connection to the grid. We assume consumers aim to {\color{black}maximize their energy benefits} by adjusting consumption behind the meter, based on the prices they face, and they may be aided in this respect by algorithms or energy service companies. The activities behind the meter are unobserved by the SO. The specific details  are outlined below:

\begin{figure}[t]
    \centering
    
    % Subfigure a
    \begin{minipage}[b]{0.4\textwidth}
        \centering
        \begin{tikzpicture}
            % Observable
    \draw[fill=blue!10] (0,1) rectangle (1,2) 
       node[anchor=west] at (-0.5, 2.3) {\small Observable};
            % Baseload
            \draw[fill=blue!20] (0,0) rectangle (2,1) node[pos=.5] {\small $d^{bl}$};
            % Sheddable
            \draw[fill=blue!30] (2,0) rectangle (3,1) node[pos=.5] {\small $d^{sd}$};
            % Arrow and Shifted
            \draw[<-, thick] (1,1.5) -- (3,1.5) node[midway, above] {\small $d^{sf, -}$};
        \end{tikzpicture}
        \subcaption{Downward shifted case}
    \end{minipage}
    \hfill
    % Subfigure b
    \begin{minipage}[b]{0.4\textwidth}
        \centering
        \begin{tikzpicture}
            % Observable
            \draw[fill=blue!10] (0,1) rectangle (4,2) node[pos=.5] {\small Observable};
            % Baseload
            \draw[fill=blue!20] (0,0) rectangle (2,1) node[pos=.5] {\small $d^{bl}$};
            % Sheddable
            \draw[fill=blue!30] (2,0) rectangle (2.5,1) node[pos=.5] {\small $d^{sd}$};
            % Arrow and Shifted
            \draw[->, thick] (2.5,0.5) -- (4,0.5) node[midway, above] {\small $d^{sf, +}$};
        \end{tikzpicture}
        \subcaption{Upward shifted case}
    \end{minipage}
    
    \caption{Decomposition of the observable load (by the SO) into its components: baseload demand $d^{bl}$, downward shifted demand $d^{sf,-}$, upward shifted demand $d^{sf,+}$ and sheddable demand $d^{sd}$.} \label{fig:dibujo_scheme_load}
\end{figure}
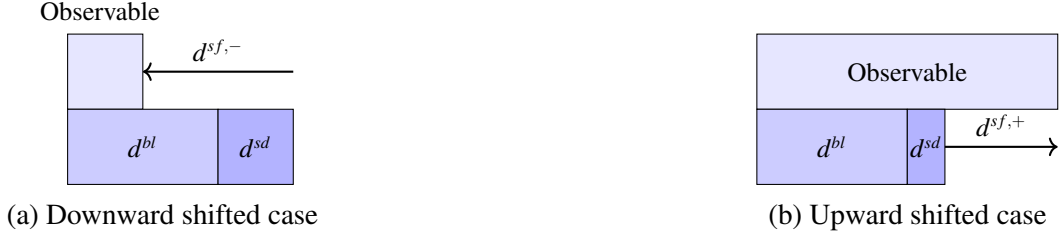

\subsection{Consumer}  In its most general form, we assume that the consumer has baseload (inflexible) demand, some self-generation (such as rooftop solar PV panels), and some flexible demand, including storage or electric vehicle charging. The demand components are illustrated in Figure~\ref{fig:dibujo_scheme_load} and detailed below:

\noindent{\underline{Net demand:} The self-generation of the consumer at any time $t$ is exogenous and random, denoted by $\hat{g}_t$. The electricity demand of the consumer is first served by this self-generation, if available. The rest of the demand, after taking the self-consumption and including both baseload and flexible components, is the net demand. The net demand is what the SO can observe and measure at the meter point. 

\noindent{\underline{Baseload demand:} This component of the net demand refers to the electricity demand from the consumer which is unresponsive to price incentives. We denote this at time $t$,  by $d^{bl}_{t}$. This has the connotation of being the necessary demand that the consumer requires at each point in time at any price below the retail market price cap, although in practice it is an empirical estimate of inflexibility across the range of prices as observed in the data.

\noindent{\underline{Shiftable demand:} The consumer can shift (advance or delay) some of its consumption when there are price incentives to do so. This can be done through adjusting the use of appliances such as washing machines, or interrupting the periodic requirements for heating and cooling, or making opportunistic use of battery storage and EV charging/discharging. We denote by $d^{sf}_{t}$, the shifted part of the net demand at any time $t$. Note that the shifted demand at any time $t$ could be a positive or negative, respectively, denoted by $d^{sf,+}_t$ and $d^{sf,-}_t$, where $d^{sf}_{t} = d^{sf,+}_t - d^{sf,-}_t$, and $d^{sf,+}_t \geq 0, d^{sf,-}_t \geq 0$. The consumer receives linear price benefits  $p^{sf,+}_t$, $p^{sf,-}_t$ per unit of energy shifted, in the positive and negative directions, respectively. However, the consumer also incurs a total cost of $c^{sf,+} (d^{sf,+}_t)^2$ and  $c^{sf,-} (d^{sf,-}_t)^2$, respectively, for shifting in the increasing and decreasing directions at any time $t$. This cost would, for example, relate to the comfort or amenity loss to the consumer through this shift, and the quadratic assumption is intended to reflect that it is likely to be an increasing function of scale. Finally, we assume that the load shifting in the positive and negative directions are {\color{black}both upper bounded by the maximum flexibility, denoted by $\overline{d}^{flex}_{t}$, and referred to as `flexibility envelope'.} 

\noindent{\underline{Sheddable demand:} This part of the net demand, denoted by $d^{sd}_{t}$, refers to discretionary consumption {\color{black}that  is still required and can be entirely or partially avoided when there is a price  incentive. We denote by  $d^{sd,-}_t$ the amount of sheddable demand that is taken. Thus,  we have by definition  $d^{sd}_t = \overline{d}^{flex}_t - d^{sd,-}_t$.}  As with shifting, the consumer receives remuneration of $p^{sd}_t$ per unit of energy shedding but also incurs a total cost of $c^{sd} (d^{sd,-}_t)^2$ for shedding demand.

\subsection{Service provider (SO)} 

The consumer needs to interact with an organization for energy delivery and billing. This organization, in different jurisdictions or applications, could take a different form such as an energy retailer, an aggregator, or a network operator. We abstract away from the specific nature of this organization for the purpose of this study and we call it simply a service provide (SO). What is shared among all of these distinct forms of `SOs' is that they need to have an accurate understanding and characterization of the demand components of the consumer. To be more precise, the SO can observe the net demand, but not necessarily the building blocks of the demand (i.e., sheddable, shifted, and baseload demands). Figure~\ref{fig:dibujo_scheme_load} illustrates the `observable' net demand by the SO, as well as the building blocks of the net demand that may be hidden from the SO. Gaining a deeper understanding of these building blocks of the net demand, without direct measurements, would benefit the SO in many ways as indicated in the Introduction. The open question is whether and how the SO can hypothesize the building blocks of net consumption and subsequently use it for better prediction. In this work, we show how to achieve this by exploiting some exogenous information in the grid (weather information, contextual grid data, voltage, etc) in addition to the historical values of the observable net demand.

\begin{table}[t]

\begin{tabular}{ccccc}
\hline
\textbf{} & \textbf{Perfectly given} & \textbf{Stats given} & \textbf{Objective} & \textbf{Decision variables} \\ \hline
    \textbf{Consumer} & $\mathbf{c},\mathbf{z},  \mathbf{u}$ & $\hat{\mathbf{g}},\hat{\mathbf{d}}$ & \color{black} {Benefit maximization} & $\boldsymbol{\theta}$  \\ \hline
\textbf{SO} & $\mathbf{z}, \overline{M}$ & $ \hat{\mathbf{g}},\hat{\mathbf{d}},\hat{\boldsymbol{\xi}}_{t}$&  Forecast error minimization & $\boldsymbol{\theta},\mathbf{u}$\\ \hline
\end{tabular}

\caption{Describing SO's and consumer's problems}
\label{table:goals_cons_SO} 
\end{table}

\subsection{Consumer and SO problem formulations}

Using the above definitions and setup, we proceed with the consumer's and SO's problem formulations. To do so, we also introduce three further notations. We define the array of `demand attributes' as {\color{black}$\mathbf{u}:= (d^{bl}_{t}, \overline{d}^{flex}_{t})_{t\in [T]}$} and the array of `costs' as $\mathbf{c}=(c^{sf,+}_t, c^{sf,-}_t, c^{sd}_t)_{t\in[T]}$ and the array of `prices' as $\mathbf{z}=(p_t, p^{sf,+}_t, p^{sf,-}_t, p^{sd}_t)_{t\in[T]}$. 

\subsubsection{Consumer's {\color{black}{benefit}} maximization}
We assume that the consumer aims to {\color{black}{maximize its overall benefit} for the entire review period $[T]$. Note that here consumer's benefit refers to the monetary compensation they receive by providing flexibility.} By review period, we mean the forward horizon over which it is seeking to optimize consumption. 
In this {\color{black}{benefit maximization}}, the consumer is assumed to know its cost parameters ($\mathbf{c}$), prices ($\mathbf{z}$), its flexibility potentials and baseload demand ($\mathbf{u}$),   the historical values of its net demand ($\hat{\mathbf{d}}$),  and its self-generation ($\hat{\mathbf{g}}$). With this information, the consumer decides how to activate its flexibility (i.e., shifting and shedding) to maximize the {\color{black}{benefit}}. We define $\boldsymbol{\theta} := (\theta_t)_{t\in [T]}=(d^{sf,+}_t,d^{sf,-}_t,d^{sd,-}_t)_{t\in [T]}$ as the vector that represents the flexibility decision variables of the consumer. With this notation, we can express the consumer problem as 
 \begin{align}\label{consumer_concrete}
\displaystyle \max_{ \boldsymbol{\theta} \in \Theta} &\; \Bigl(\ell(\boldsymbol{\theta}) := \sum_{t\in [T]}[Q_t(\theta_t) + L_t(\theta_t)] \Bigr) &
 \end{align}
where $\ell$ is consumer's utility function and $\Theta$ is the flexibility feasibility set both explained below. The utility function $\ell$ defined in Eq.~\eqref{consumer_concrete}  is the {\color{black}{benefit} comprising of a quadratic cost and a linear term}, respectively, denoted by $Q_t$ and $L_t$. The quadratic part represents the total comfort cost of demand shifting and shedding,  $Q_t(\theta_t):=-c^{sf,+} (d^{sf,+}_t)^2-  c^{sf,-} (d^{sf,-}_t)^2
-c^{sd} (d^{sd,-}_t)^2$.  The linear part
represents the financial exchanges with the SO, which is the difference between the demand flexibility gain and the consumption cost, {\color{black}$L_t(\theta_t):=p^{sf,+}_{t}d^{sf,+}_{t}+p^{sf,-}_{t}d^-_{t} +p^{sd}_{t} d^{sd,-}_{t} -p_{t}(d^{bl}_{t} 
     + d^{sf,+}_{t}-d^{sf,-}_{t}+\overline{d}^{flex}_t-d^{sd,-}_{t} -\hat{g}_{t} )]$}.

To understand the feasible region in which the consumer chooses its optimal flexibility decision, we define the \emph{flexibility feasibility set}, $\Theta$, which comprises the set of possible actions of the consumer for its flexible (shifted and sheddable) behavior. Essentially, $\Theta$ incorporates  {\color{black} the flexibility envelope that restricts physically the maximum level of flexibility at time  $t$. 
We, therefore, represent the flexibility feasibility set $\Theta$ as $\Theta(\overline{d}^{flex}_{t})  $ to make explicit the dependence with respect to the flexibility envelope}. Given the envelope, the feasibility set is expressed as follows:
\begin{equation}\label{Theta_set}
\begin{aligned}
\Theta:= 
&\left\{
\begin{array}{ll}
0 \leq d^{sd,-}_{t} \leq \overline{d}^{flex}_{t} & \forall t \in [T] \quad 
\\
0 \leq d^{sf,+}_{t} \leq  \overline{d}^{flex}_{t}  &\forall t \in [T] \quad  \\
0 \leq d^{sf,-}_{t} \leq  \overline{d}^{flex}_{t} &\forall t \in [T] \quad  \\
\displaystyle\sum_{t\in [T]}d^{sf,+}_{t}=\sum_{t\in [T]}d^{sf,-}_{t} &  \\
\end{array}
\right.
\end{aligned}
\end{equation}
The first three lines in the feasibility set described above ensure that the shedding, positive shifting, and negative shifting, all satisfy the flexibility envelope.  The fourth line ensures that the overall demand shifting is energy neutral and does not carry over to the next review cycle; i.e., any shifting in the positive/negative direction must be reversed before the review period is over.

\subsubsection{SO's forecast error minimization} 

The ultimate goal of the SO is to minimize the (out-of-sample) forecast error for net demand and, in the process,  to estimate the consumer's unobserved demand components, including the flexibility responses/decisions, $\boldsymbol{\theta}$, and  baseload demand and flexibility potentials, 
$\mathbf{u}$. We assume that the SO knows the array of price signals $\mathbf{z}$ and the maximum load capacity $\overline{M}$. It is generally the case that the SO has information about self generation $\hat{\mathbf{g}}$ (as these renewable facilities are usually metered separately for subsidies and/or regulatory requirements), past observed net demand measurements $\hat{\mathbf{d}}$ and various relevant weather/grid data $\hat{\boldsymbol{\xi}}_t$ for the entire review period $[T]$.
Table \ref{table:goals_cons_SO} summarizes the problem setting.
\section{Solution using inverse optimization (IO)}\label{sec:IO}

In this section, we first introduce the IO methodology and then explain how it can be further developed for unobservable demand response.
\subsection{Theoretical background on inverse optimization}

In the terminology of Inverse Optimization (IO),  a \emph{forward} optimization program (FOP) is proposed whose solutions are observed. This FOP is parametrized on some exogenous observations, $\mathbf{w}$, and some endogenous parameters, $\mathbf{u}$.  Hence the formulation in a deterministic setting can be expressed as:
\begin{equation}\label{def:FOP_generic}
\begin{aligned}
 \text{FOP}(\mathbf{w} | \mathbf{u}):= 
&\left\{
\begin{array}{ll}
\displaystyle\max& \quad \ell(\boldsymbol{\theta};  (\mathbf{w} | \mathbf{u})) \\
 \quad\text{s.t.}&       \quad  \boldsymbol{\theta}
            \in \Theta(\mathbf{u} )
\end{array}
, \right.
\end{aligned}
\end{equation}
where for any value of $ \mathbf{u}$, we assume that $\Theta(\mathbf{u} )$ is a closed convex set and  $\ell$ a differentiable and concave utility function with respect to $\boldsymbol{\theta}$. In a non-deterministic setting (such as our problem), the IO solution must consider the observation noise, such that the IO model becomes a supervised learning problem with multivariate output. More formally, therefore, the training dataset, $\mathcal{D}$, is built up by $S$ pairs of observations $\mathcal{D}_{S}:= \{ (\mathbf{w}_s, \boldsymbol{\theta}_s)\}_{s\in [S]}$ and the goal is to find $\hat{\mathbf{u}}$ that would make $\boldsymbol{\theta}_s$ ``optimal'' for the $s$-th FOP. That is,  $\text{FOP}(\mathbf{z}_s | \hat{\mathbf{u}} )$. Thus, having estimated the parameter $\mathbf{u}$ by $\hat{\mathbf{u}}$, and given a new observation $\mathbf{w}'$,  $\text{FOP}( \mathbf{w}'| \hat{\mathbf{u}} )$ is used to estimate $\boldsymbol{\theta}'$ \citep{saez2017short}.

\subsection{Applying IO to unobservable demand response}

The FOP model is naturally delivered by the consumer's problem as defined in Eq.~\eqref{consumer_concrete}.   Adapting the basic approach as described above, we proceed in detail, as follows:

\subsubsection{The forward or reconstruction program}

Given the training dataset $\mathcal{D}_{S}:= \{ ((\mathbf{c}_s,\mathbf{z}_s,\hat{\mathbf{g}}_s), \hat{\mathbf{d}}_s)\}_{s\in [S]}$ made by $S$ training samples (which may  correspond to a set of days), built on pairs of $T-$dimensional exogenous observations (cost, price and self generation data), $(\mathbf{c}_s,\mathbf{z}_s,\hat{\mathbf{g}}_s)$,  and observed load measurements $\hat{\mathbf{d}}_s$. In this setting, $(\mathbf{c}_s,\mathbf{z}_s,\hat{\mathbf{g}}_s)$ plays the role of $\mathbf{w}$ in program \eqref{def:FOP_generic}. \\
Thus, the data-driven IO model tries to find estimates of demand attributes $\hat{\mathbf{u}}_s$ of the  parameters 
$\mathbf{u}_s$, where {\color{black}$\mathbf{u}_s:=(d^{bl}_{t}, \overline{d}^{flex}_{s,t})$,}
which would make the observed $s$-th  demand $\hat{\mathbf{d}}_s$ ``optimal'' for the
 $s$-th FOP,  $\text{FOP}((\mathbf{c}_s,\mathbf{z}_s,\hat{\mathbf{g}}_s) | \mathbf{u}_s)$,
defined as follows:
\begin{equation}\label{def:FOP_concrete}
%\begin{alignedat}{2}
%\text{FOP}((\mathbf{c}_s,\mathbf{z}_s,\hat{\mathbf{g}}_s) | \mathbf{u}_s):=
%&\left\{
\begin{array}{lll}
\displaystyle \max & \ell(\boldsymbol{\theta}_s;((\mathbf{c}_s,\mathbf{z}_s, \hat{\mathbf{g}}_s)| \mathbf{u}_s))&\\
 \quad\text{s.t.}&   \boldsymbol{\theta}_s=(\mathbf{d}^{sf,+}_s,\mathbf{d}^{sf,-}_s, \mathbf{d}^{sd,-}_s) \in \Theta({\color{black}\overline{d}^{flex}_{s,t}})    \\
   &  d^{sf}_{s,t} = d^{sf,+}_{s,t} - d^{sf,-}_{s,t}& \forall t \in [T] \\
       & d^{sd}_{s,t}={\color{black}\overline{d}^{flex}_{s,t}}-d^{sd,-}_{s,t}& \forall t \in [T]\\
        %     &  \displaystyle  \sum_{t\in [T]}(\delta^+_{s,t}+ \delta^-_{s,t})\leqslant T^{\max} & \\
                    %             & \delta^+_{s,t}+ \delta^-_{s,t} \leqslant 1 &  \forall t \in [T]\\
                %    & \delta^{sf,+}_{s,t}, \delta^{sf,-}_{s,t} \in \{ 0,1 \} &  \forall t \in [T] 
\end{array}
%\end{alignedat}
\end{equation}

%where the function $\ell$ is the utility function defined by eq.\eqref{utility_function}
 % is evaluated on the decision variables  $\boldsymbol{\theta}_s$.
  %but we write also its dependence on
%  the  exogenous parameters $\mathbf{z}_s:=(\mathbf{c}_s^+,\mathbf{c}^{-}_s, \mathbf{c}^{0}_{s},\mathbf{p}_s^+,\mathbf{p}^{-}_s, \mathbf{p}^{0}_{s}, \mathbf{p}_s, \mathbf{g}_s)$  built on cost and price values, $\mathbf{c}_s^+,\mathbf{c}^{-}_s, \mathbf{c}^{0}_{s}, \mathbf{p}_s^+,\mathbf{p}^{-}_s, \mathbf{p}^{0}_{s}, \mathbf{p}_s$ and the exogenous local generation, $\mathbf{g}_s$,

Having introduced the end-users' FOP, next we explain how the proposed framework works: (i) a \emph{learning step} where the data-driven IO model is fitted by using a training dataset, and (ii) a \emph{forecasting step} where the  time-series forecast output is computed.

{\color{black}The flexibility envelope  $\overline{d}^{flex}_{s,t}$, which is constrained by the input parameter $\overline{M}$, 
(in the numerical experiments is equal to  the maximum observed consumption $\max_{s \in [S], t \in [T]}|\widehat{d}_{s,t}|$),  is a random function itself because it may evolve over time from day to day, and hence, should be estimated from past data. To do so, we follow the approach introduced by \cite{saez2017short}, which proposes an affine decision rule  to capture the dependence  between regressors and power:
\begin{align}
    \overline{d}^{flex}_{s,t}& = \beta^{flex}_{0}+
    \langle \boldsymbol{\beta}^{flex} , \hat{\boldsymbol{\xi}}_{s,t} \rangle
    \label{linear_rule_plus}
\end{align}
 Moreover, $\hat{\boldsymbol{\xi}}_{s,t}$ is the vector of external features  (for example weather information or contextual grid data, etc) corresponding to day $s$ at hour $t$ of the inverse optimization program used to estimate
the maximum availability of flexible demand at time $t$ on day $s$.  In order to get the recovery of the IO parameters (via Theorem 4 in \cite{aswani2018inverse}) and mitigate the  over-fitting effect  of the flexibility envelope,
we need to restrict the parameter space to a reduced class, namely the one given by the following equation:

\begin{equation}
    (\mathbf{d}^{bl},[\beta^{flex}_0,\boldsymbol{\beta}^{flex}]) \in [0,\overline{M}]^T\times \Lambda_{\beta} 
\end{equation} where 
  the set 
$\Lambda_{\beta}$ is defined by the following equation:
\begin{equation}\label{paremeter_space_eqs }
\Lambda_{\beta} := \left\{ [\beta^{flex}_0, \boldsymbol{\beta}^{flex}] : 
\begin{aligned}
    & |\beta^{flex}_{0}| \leqslant M_0, \\
    & 0 \leqslant \beta^{flex}_{0} + \langle \boldsymbol{\beta}^{flex}, \boldsymbol{\xi} \rangle \leqslant \overline{M}, \quad \forall \boldsymbol{\xi} \in \Xi
\end{aligned} 
\right\}
\end{equation}
being  $\overline{M}_0$ bounds the intercept,
and the robust constraint \eqref{paremeter_space_eqs } is added to ensure  that the flexibility envelope value does not surpass the maximum load capacity $\overline{M}$ for all possible realizations of the regressors and not only for the ones observed in the past as suggested by \cite{saez2016}. The robust constraint in \eqref{paremeter_space_eqs } parameterized on the uncertainty set $\Xi$ (which encodes the range of possible values of the regressors) can be finitely reformulated
by introducing auxiliary dual variables as done in
\cite{saez2016}.

 }
\subsubsection{Learning}

Having observed a  time series given by set of  $[S]$ days of the  demand per time period $t$, $\{(\widehat{d}_{s,t})_{t\in [T]}\}_{s\in [S]}$, which form the training dataset, the learning step  is given by the solution of the following data-driven inverse optimization program:

\begin{subequations}  
\begin{align}
  \min &  \displaystyle \sum_{s \in [S]}\omega_s\left\|\left( [\mathbf{d}^{bl}+\mathbf{d}^{sf}_{s} + \mathbf{d}^{sd}_{s}-\hat{\mathbf{g}}_{s}] -\mathbf{\widehat{d}}_{s}\right)_s \right\|^2_2  + \lambda\left\|\boldsymbol{\beta}^{flex} \right\|^2_2 \label{obj_function:learning}\\
 \text{ \text{s.t.} }   
 & d^{sf}_{s,t} = d^{sf,+}_{s,t} - d^{sf,-}_{s,t}, \forall s \in [S], \forall t \in [T] \label{eq:elastic}\\
                 &  d^{sd}_{s,t}= \beta^{flex}_{0}+
    \langle \boldsymbol{\beta}^{flex} , \hat{\boldsymbol{\xi}}_{s,t} \rangle-d^{sd,-}_{s,t}, \forall s \in [S],\forall t \in [T]   \label{eq:sheddable}\\
                   & (\mathbf{d}^{bl},[\beta^{flex}_0,\boldsymbol{\beta}^{flex}]) \in [0,\overline{M}]^T\times \Lambda_{\beta} \label{eq:compact_feas_region}\\
                             & \boldsymbol{\theta}_s:= (\mathbf{d}^{sf,+},\mathbf{d}^{sf,-}, \mathbf{d}^{sd,-})_s, \;\mathbf{u}_s=(d^{bl}_{t}, \beta^{flex}_{0}+
    \langle \boldsymbol{\beta}^{flex} , \hat{\boldsymbol{\xi}}_{s,t} \rangle)_s,  \forall   s \in [S]\\
  &   
\boldsymbol{\theta}_s \in \arg \max    \ell(\boldsymbol{\theta}_s;( (\mathbf{c}_s,\mathbf{z}_s, \hat{\mathbf{g}}_s )| \mathbf{u}_s)) \label{lower_level1} \\
       &    \hspace{0.5cm}   \text{ \text{s.t.} }  \displaystyle\sum_{t\in [T]}d^{sf,+}_{s,t}=\sum_{t\in [T]}d^{sf,-}_{s,t} :   (\kappa_{s})    \\
                        &  \hspace{0.8cm}   d^{sf,+}_{s,t}\leq  \beta^{flex}_{0}+
    \langle \boldsymbol{\beta}^{flex} , \hat{\boldsymbol{\xi}}_{s,t} \rangle   :   (\mu^+_{s,t } \geqslant 0), \forall t \in [T]\\
                    & \hspace{0.8cm}    d^{sf,-}_{s,t}\leq  \beta^{flex}_{0}+
    \langle \boldsymbol{\beta}^{flex} , \hat{\boldsymbol{\xi}}_{s,t} \rangle  :  (\mu^-_{s,t }\geqslant 0),  \forall t \in [T]\\
                 & \hspace{0.8cm}    d^{sd,-}_{s,t}\leq \beta^{flex}_{0}+
    \langle \boldsymbol{\beta}^{flex} , \hat{\boldsymbol{\xi}}_{s,t} \rangle  \quad : \quad  (\mu^0_{s,t }\geqslant 0) ,  \forall t \in [T]\\
                 & \hspace{0.8cm}0 \leq d^{sf,+}_{s,t}  :  (\nu^+_{s,t }\geqslant 0) ,  \forall t \in [T]\\
                                  & \hspace{0.8cm} 0 \leq d^{sf,-}_{s,t}  :  (\nu^-_{s,t }\geqslant 0) ,  \forall t \in [T]\\
                                                   & \hspace{0.8cm}0 \leq d^{sd,-}_{s,t}  :   (\nu^0_{s,t }\geqslant 0) ,  \forall t \in [T]  \label{eq:end_learning}
\end{align}
\end{subequations}  
{\color{black}where $\left\| \cdot \right\|_2$, denotes the $2-$norm,
$\omega_s$ is the weight of the forecast error on day $s\in [S]$ \citep{saez2016} and the variables in parenthesis in the constraints of the $s-$th lower level problem denote the respective dual variables.
We set $\omega_s:=\left(\frac{s}{S}\right)^\alpha/(\sum_{s'\in [S]}\left(\frac{s'}{S}\right)^\alpha)$ the normalized weights of the training days, with $ \alpha \geqslant 0$ the  forgetting factor  parameter to represent discounting older data. If $\alpha=0$, then  all observations are equally weighted and when $\alpha$ increases, more weight is given to recent observations, as suggested by \cite{saez2016,lu2018data}. Also, we add the term  $\lambda\left\|\boldsymbol{\beta}^{flex} \right\|^2_2$ in \eqref{obj_function:learning} (as done in Ridge regression)
by using  a regularization parameter $\lambda > 0$.  Note that the affine decision rule parameters, $[\beta^{flex}_0,\boldsymbol{\beta}^{flex}]$,  are determined by the IO program, as they are treated as decision variables within the upper-level model.
}

\subsection{Component estimation and forecasting}\label{section_estimation_forecasting_two_step}
{\color{black}Having introduced the parameterizations in terms of the affine decision rules and the decision variables for the flexibility envelopes, we  compute $  \widehat{\boldsymbol{\beta}}$ and
 $\widehat{\mathbf{d}}^{bl}$ 
being  the optimal solutions of  $\boldsymbol{\beta}:=[\beta^{flex}_{0},\boldsymbol{\beta}^{flex} ]$ and $\mathbf{d}^{bl}$ 
in the problem defined by \eqref{obj_function:learning}-\eqref{eq:end_learning}, respectively. Thus, we can compute $\widehat{\mathbf{u}}:=(\widehat{\mathbf{d}}^{bl},\overline
{d}^{flex}_{S+1,t})$ where
\begin{align}
   \overline
{d}^{flex}_{S+1,t}& = \hat{\beta}^{flex}_{0}+   \langle \hat{\boldsymbol{\beta}}^{flex} ,  \hat{\boldsymbol{\xi}}_{S+1,t} \rangle .
\end{align}
Having computed $\hat{\mathbf{u}}$, the forecast output for the day $S+1$ is given by solving the following forward program given the exogenous observations for the day $S+1$, that is cost, price and local generation  data on day $S+1$, $(\mathbf{c}_{S+1},\mathbf{z}_{S+1}, \hat{\mathbf{g}}_{S+1})$:
\[
\begin{array}{c@{\,}l@{~}l}\max &  \displaystyle \ell(\boldsymbol{\theta}_{S+1}; ((\mathbf{c}_{S+1},\mathbf{z}_{S+1}, \hat{\mathbf{g}}_{S+1}) | \hat{\mathbf{u}}))\\
           \text{s.t.}          &       d^{sf}_{S+1,t}=d^{sf,+}_{S+1,t}-d^{sf,-}_{S+1,t}& \forall t \in [T]\\
       & d^{sd}_{S+1,k,t}=\overline
{d}^{flex}_{S+1,t}-d^{sd,-}_{S+1,t}& \forall t \in [T]\\
      & \boldsymbol{\theta}_{S+1}=(\mathbf{d}^{sf,+}, \mathbf{d}^{sf,-}, \mathbf{d}^{sd,-})_{S+1}  &\\
                  & \boldsymbol{\theta}_{S+1} \in \Theta( \overline
{d}^{flex}_{S+1,t}) .&
\end{array}
\]

After solving the above program and getting $(\mathbf{d}^{sf,+}, \mathbf{d}^{sf,-}, \mathbf{d}^{sd,-})_{S+1}$, we can compute the forecast net load per time $t$ for day $S+1$, $\hat{d}_{S+1,t}$,  as the sum of the (net) baseload demand, $\hat{d}^{bl}_{t}-\hat{g}_{S+1,t}$, plus the flexible demand, $d^{flex}_{S+1,t}:=d^{sf}_{S+1,t}+d^{sd}_{S+1,t}$ as follows $
    \hat{d}_{S+1,t}:=(\hat{d}^{bl}_{t}-\hat{g}_{S+1,t})+d^{flex}_{S+1,t}$.
The Appendix \ref{sec:reformulation}  develops a single-level tractable reformulation by exploiting the Karush-Kuhn-Tucker (KKT) conditions for the lower-level and the underlying conic structure of the data-driven IO problem.
}

\section{Empirical research }\label{sec:case_study}

Using the above IO methodology,  the logical flow of the empirical research questions that we address are as follows:

{\color{black}

\textbf{Q1: Using net demand with structural assumptions and a presumed optimization model for the flexible and baseload components of domestic demand and their price responses, how well is the above IO formulation able to estimate the latent components of flexible and baseload demand?} This will establish proof of concept and computational tractability.

\textbf{Q2: Can the resulting latent component estimates from IO  be accurately predicted out of sample, using market-wide information such as weather and periodic effects, but without assuming knowledge of the appliance activities?} The predictability of the components needs to be established against benchmark statistical and machine learning methods. 

\textbf{Q3: Are forecasts of net demand obtained by summing the latent components more accurate than forecasts of the aggregate net demand series?} This is needed to validate the value of the decomposed approach for forecasting. 

\textbf{Q4: With data that provides explicit knowledge of the domestic appliance activities, do these latent components from IO have a credible relationship to the appliance activities and thereby appear valid?}  Plausible statistical relationships to the data will establish construct validity.

\textbf{Q5: Overall, can the price response of domestic customers be more accurately estimated based upon an IO decomposition of flexible and baseload components than by conventional benchmark methods?}  This is needed to justify the overall proposition that IO can be useful in practice compared to benchmark forecasting methods.

We progress the analysis of Q1-Q5 through a set of smart household data, complied by  
\cite{datakaggle}
and openly available on the Kaggle platform to facilitate replication studies. It contains the smart meter readings in kW from a US household of all of the main household appliances at minute-by-minute granularity for  350 days of house appliances, together with the weather conditions of that particular region. It does not contain any retail price data. We do however know the typical retail prices for that region (Ohio) in 2018. This example replicates the perspective of a network operator who would not know the retailer-customer tariff details precisely.  We use this to test the application of the IO approach and sense-check the identified components of flexible and inflexible components against actual appliance use. 

%The forecasting questions are also evaluated on this data, but for Q5, which relies upon precise TOU  responses, we use consumer data from Japan, made available by \cite{kiguchi2021predicting}. This is intraday net load data from a sample of customers responding to price signals for several months in 2018. This example therefore replicates the other case of a retailer setting TOU tariffs, observing the metered net loads but not knowing exactly how the appliances were being used. The logic of our analysis is, therefore, that having established the construct validity of the component estimates on the first data set, we are justified in applying it in the second data set in order to compare its forecasting performance under TOU prices against conventional benchmarks.

The IO method is benchmarked against two methods: a time-series model with exogenous variables (SARIMAX), and also against a non-linear, machine-learning method,  XGboost \citep{chen2016xgboost}.  The performance metrics  used for evaluations of the forecasts are the usual  mean absolute error (MAE) and the  root mean square error (RMSE) defined on the Appendix \ref{performance_metrics}.

\subsection{Performance effectiveness and construct validity}

The smart household data from \cite{datakaggle} is used to test the empirical performance of the proposed  IO method against conventional benchmarks. It also allows us to check that the identified flexible and baseload components do in fact represent what they are intended to model, i.e., that they have construct validity. We do that by regressing the estimated flexible and baseload component activities on the known usage of household appliances as well as autoregressive lags. We expect the discretionary use of appliances to feature more strongly in the flexible component but less so in the baseload component, whilst the autoregressive effects would relate to adaptation and habit and should feature more strongly in the baseload component.  The data measurements include the kitchen appliances (e.g. fridge, dishwasher, microwave), various rooms (e.g. kitchen, living room, home office), furnace, well, barn, garage door, etc., as well as solar generation and weather conditions. The minute-by-minute data is from a sample of US households in Ohio taken over 2018 (as also analyzed in \cite{tlenshiyeva2024data}).  We restrict the analysis to weekdays.   Recall that the net load is consumed energy minus the solar generation.

Even though we do not know the precise tariff details for the customer, we expect the IO method to be robust to noisy assumptions and we introduce a typical tariff for that location at that time (based upon inspection of published rates from various local utilities) available at Appendix \ref{data_IO_ohio_tokyo}.  
The training data uses two months, weekdays only, and the goal is to predict the first five days of the next month. We consider two seasonal sets of results.  

We  show the summary results in Tables \ref{tab:SEASON1}-\ref{tab:SEASON3}.  Evidently, the IO is more accurate than the benchmark methods for forecasting.   Figure~\ref{Results_case_net_load} shows the net demand (demand is referred to as load in the Kaggle data source)  for the two cases (upper  January to March, and  lower  July to September) under the optimal hyperparameter tuning via grid-search. Similarly, Figures \ref{Results_case_net_baseload_load} and \ref{Results_case_flex_load} show the net baseload demand and flexible profiles estimated as part of the modeling. 

In all figures, the peak price periods are shown shaded in pink. Whilst the effect of peak pricing is not evident in the net load profiles shown in Figure~\ref{Results_case_net_load}, nor in the baseload components shown in Figure~\ref{Results_case_net_baseload_load}, it is very evident in Figure~\ref{Results_case_flex_load}. In Figure~ \ref{Results_case_flex_load}, we can see  that only the IO approach is showing a distinct reduction in the flexible load component during the hours of peak pricing. This is, of course, what we would hope to see, as the IO approach is designed to identify this component. Visually, it is a very reassuring display that the IO is indeed providing a new component that the benchmark methods cannot reveal.

Regarding relative accuracy, according to Tables~\ref{tab:SEASON1}, for Winter, IO demonstrates the lowest MAE (0.2077) and RMSE (0.2654) for the ``Net load" category, outperforming all the benchmark methods. XGBoost and SARIMAX have higher MAE and RMSE values in comparison to IO.
%but still outperform Linear in the same category. %Linear shows the highest errors across most categories. 
Tables~\ref{tab:SEASON3} shows that the results for summer exhibit slight shift. While IO maintained the lowest MAE (0.1778) and RMSE (0.2326) for ``Net load",  SARIMAX performed comparably better than XGBoost
but with higher errors than IO.

According to these figures, IO demonstrates the most accurate performance across all seasons, consistently achieving the lowest errors in the ``Net load". SARIMAX  generally follows. Surprisingly, XGBoost, despite having some strong performances in the baseload,
$d^{bl}$, and the flexible load, $d^{flex}$, often has the highest errors for ``Net load" metrics.  Most importantly for our purposes, from Figures \ref{Results_case_net_baseload_load} and \ref{Results_case_flex_load} it is evident that the baseload component is more regular and predictable than the flexible component, which is intuitive and reassuring.  Overall, these results appear to convincingly support research questions, Q1 ad Q2,  in establishing the feasibility and comparative value of the IO approach.

In relation to Q3, which focuses upon the inherent modeling element of predicting the components separately, and then summing to get the Net load forecast, Tables \ref{tab:SEASON1}-\ref{tab:SEASON3} also report some justification. The rows designated Net Load and Net Load 2 display the accuracies for predicting on the basis only of the aggregate Net Load data  versus the disaggregated approach of predicting the components and summing. 
{\color{black}
Across the two seasons, the methods show varying degrees of improvement from ``Net load'' to ``Net load 2''. XGBoost verified substantial improvements, with reductions in MAE ranging from approximately 44.3\% in Winter to 32.4\% in Summer, and reductions in RMSE from approximately 42.4\% in Winter to 33.5\% in Summer. SARIMAX exhibited consistent improvements, with MAE reductions from 42.8\% in Winter to 23.7\% in Summer, and RMSE reductions from 40.2\% in Winter to 19.2\% in Summer. Overall, all methods generally performed better with ``Net load 2,'' with XGBoost and SARIMAX showing the most significant improvements.
In other words, the principle of component-based forecasting appears to be justified.}

\begin{figure}
\centering
 \begin{minipage}[t]{0.99\textwidth}
\subcaption{Winter}{%
  \includegraphics[width=0.99\textwidth]{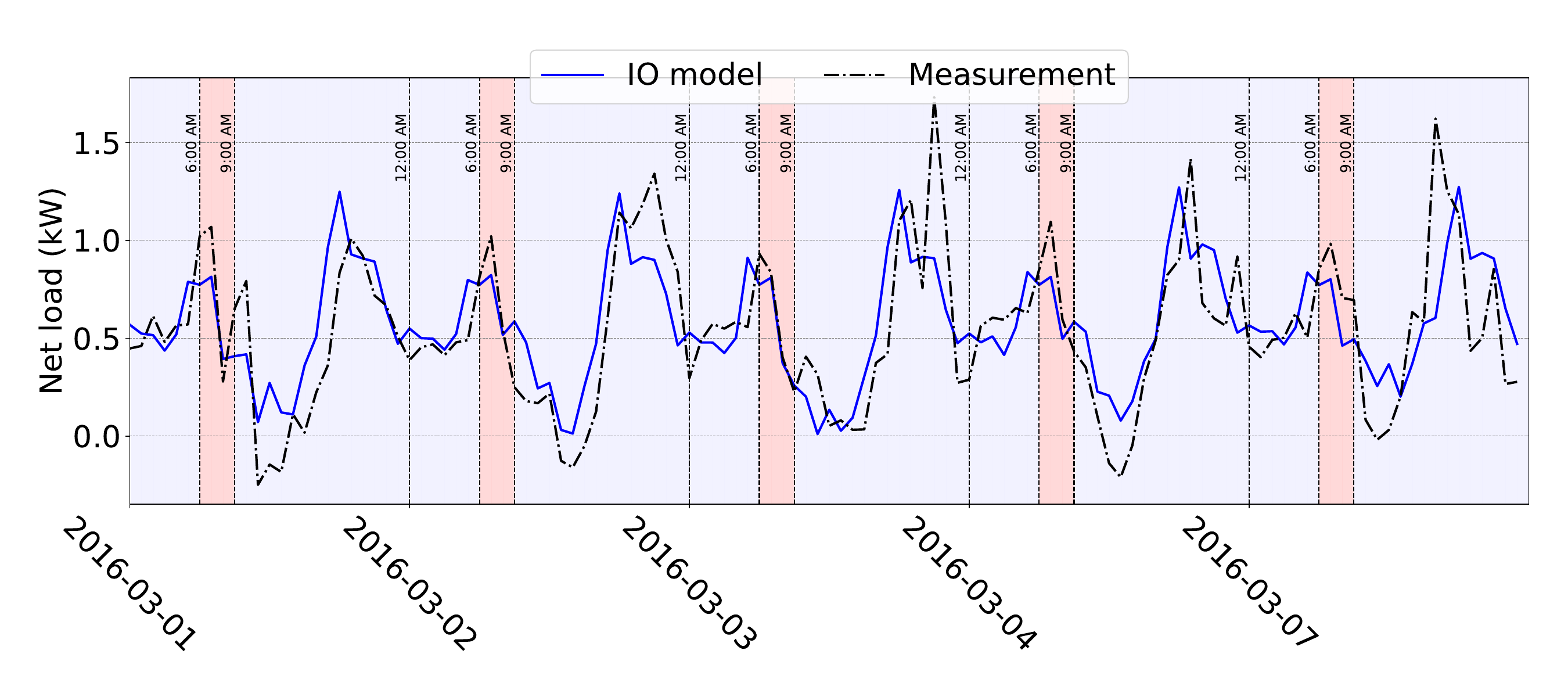}%
  \label{}
}%
\end{minipage}
%\subfloat[Season 2]{%
%  \includegraphics[width=0.5\textwidth]{figures/Fig_forecast_netload_customer_1_training_from2016-04-01_to_2016-05-31.pdf}%
%  \label{}
%}%

\centering
 \begin{minipage}[t]{0.99\textwidth}
\subcaption{Summer}{%
  \includegraphics[width=0.99\textwidth]{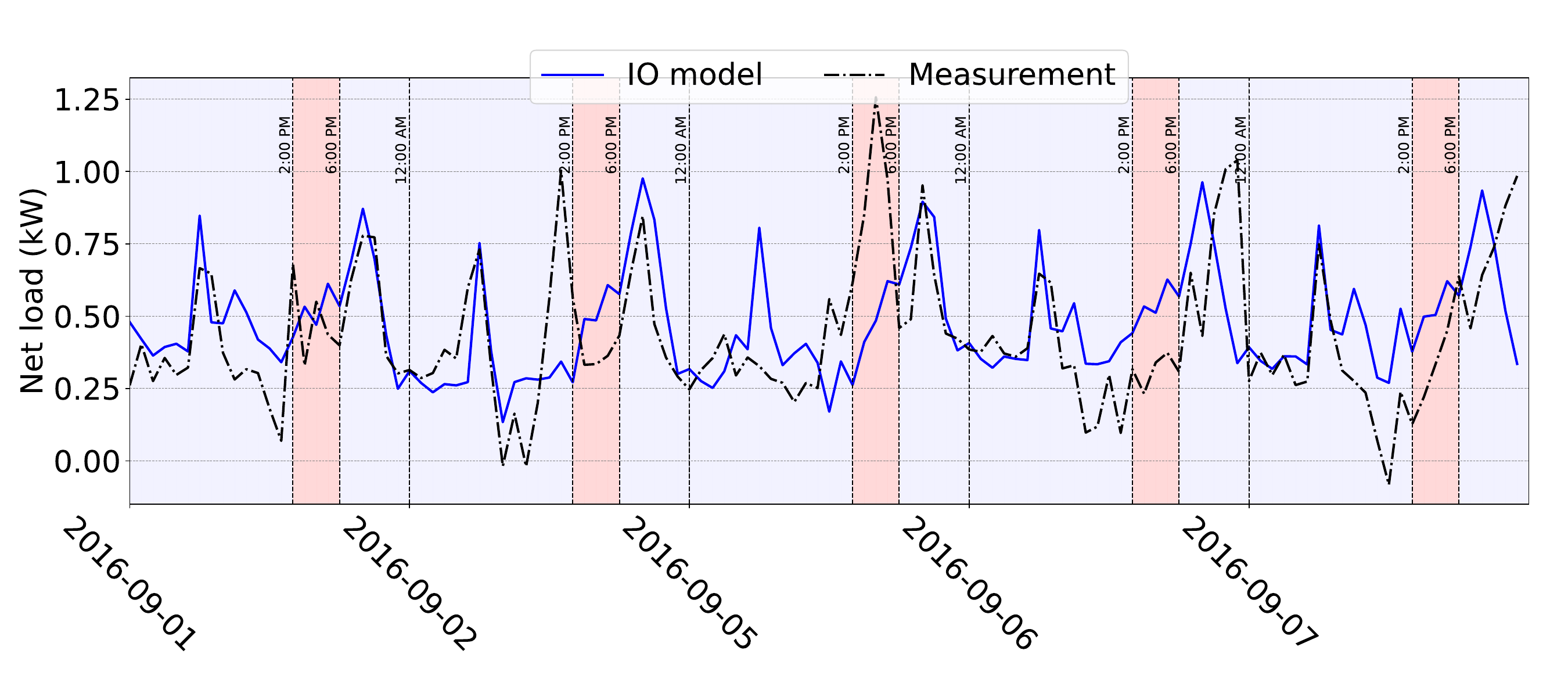}%
  \label{}
}%
\end{minipage}
%\subfloat[Season 4]{%
%  \includegraphics[width=0.5\textwidth]{figures/Fig_forecast_netload_customer_1_training_from2016-10-03_to_2016-11-30.pdf}%
%  \label{}
%}%

\caption{Net load profiles and performance error metrics  under best hyperparameter tuning}\label{Results_case_net_load}
\end{figure}

\begin{figure}
\centering
  % Primera figura (Winter)
    \begin{minipage}[t]{0.99\textwidth} % t para alinear en la parte superior
        \centering
        \subcaption{Winter} % caption sin numeración
        \includegraphics[width=\textwidth]{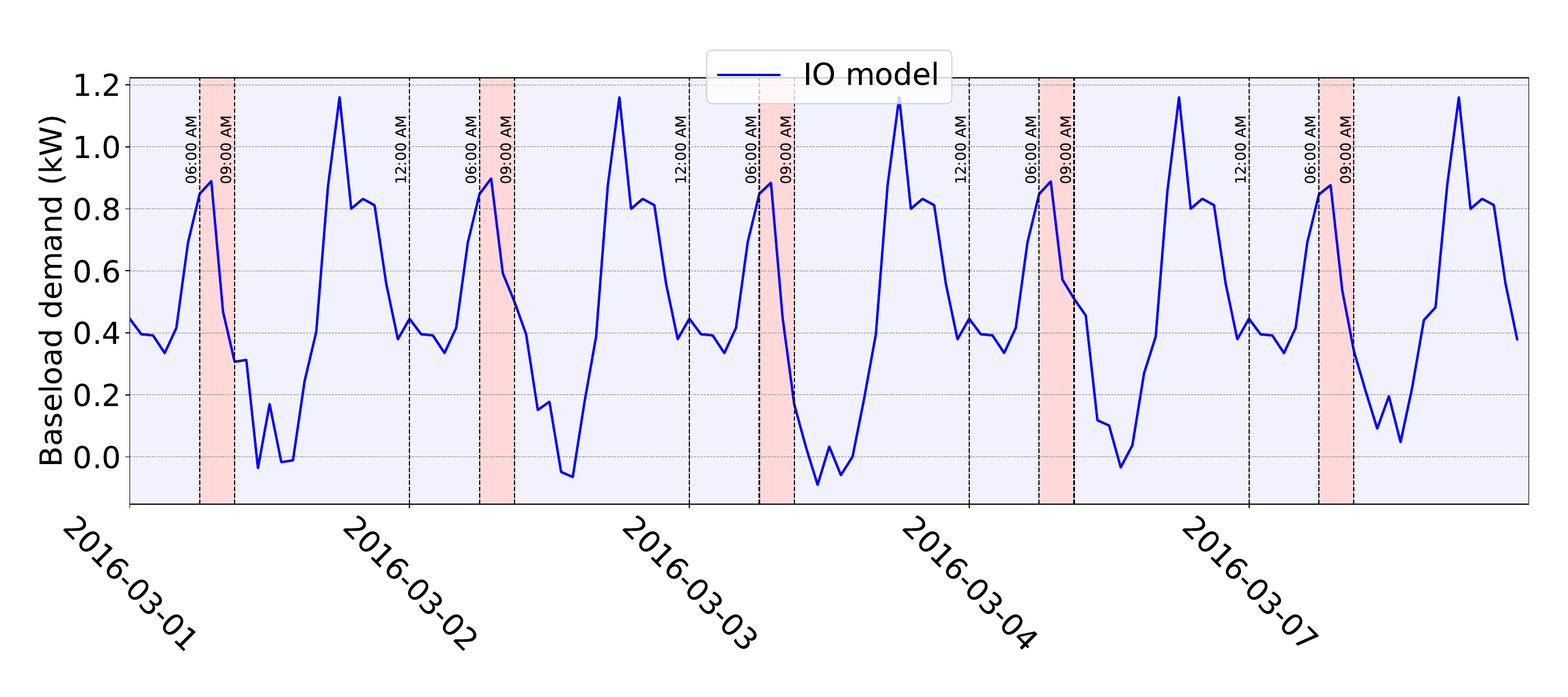}
    \end{minipage}%
    
   % \vspace{0.5cm} % Espacio vertical entre las figuras

    % Segunda figura (Summer)
    \begin{minipage}[b]{0.99\textwidth} % t para alinear en la parte superior
        \centering
        \subcaption{Summer} % caption sin numeración
        \includegraphics[width=\textwidth]{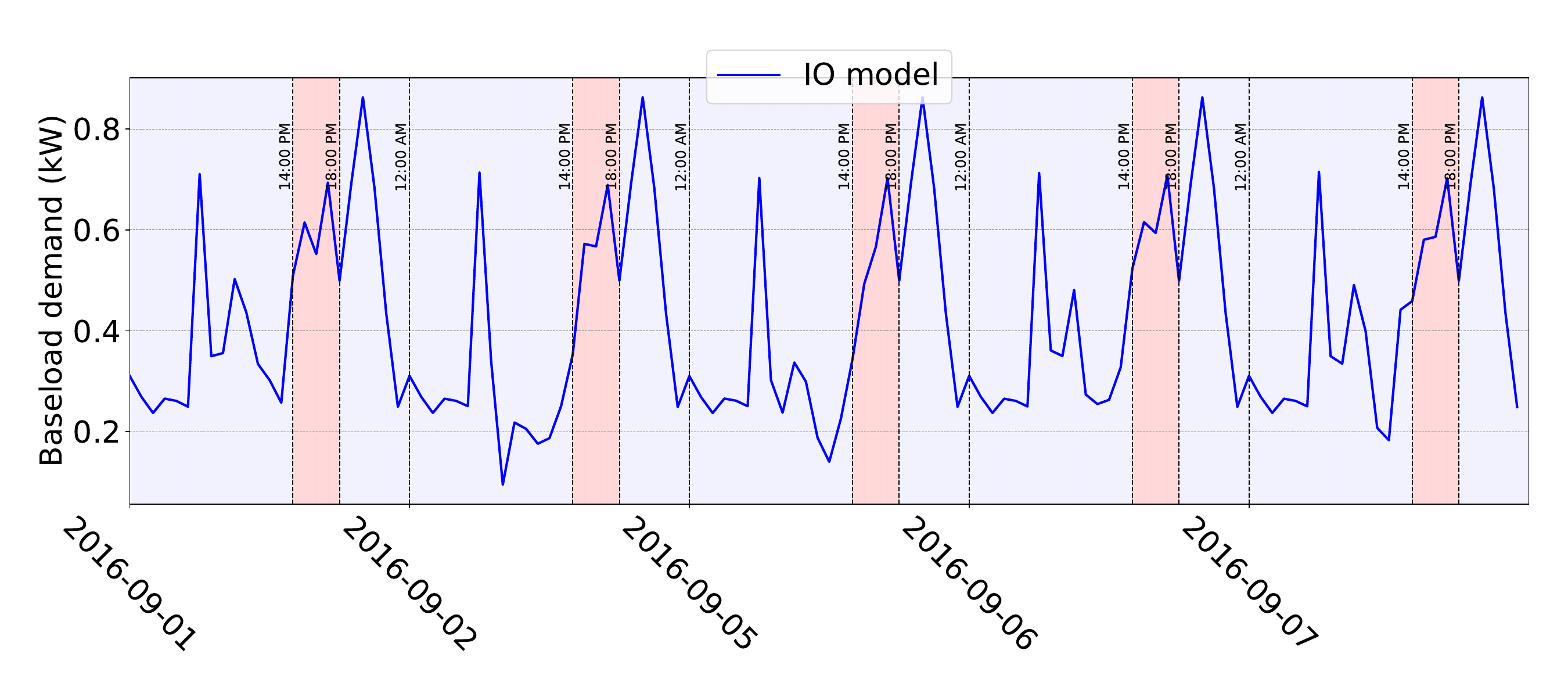}
    \end{minipage}%
%}%

\caption{Net baseload demand profiles and performance error metrics under best hyperparameter tuning}\label{Results_case_net_baseload_load}
\end{figure}

\begin{figure}
\centering
 \begin{minipage}[t]{0.99\textwidth}
\subcaption{Winter}{%
  \includegraphics[width=0.99\textwidth]{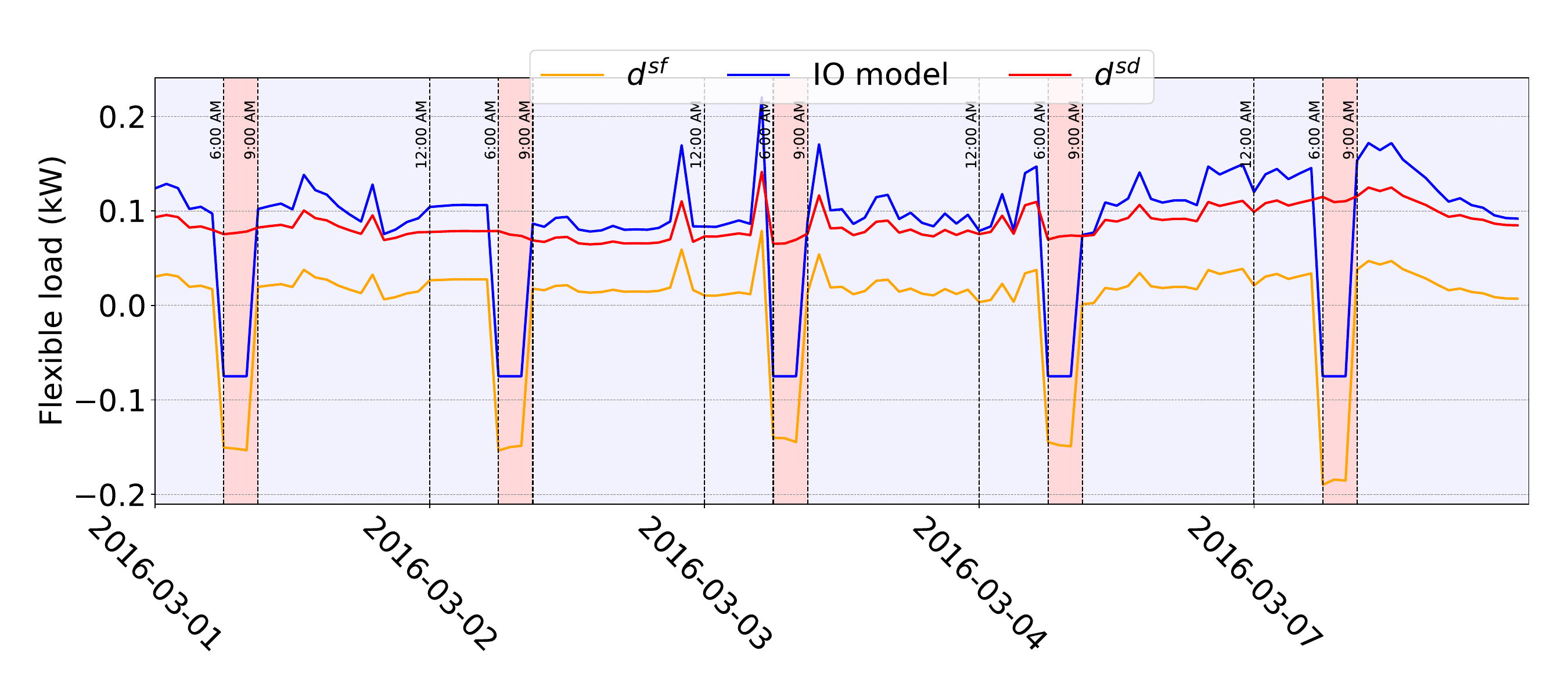}%
  \label{}
}%
\end{minipage}
%\subfloat[Season 2]{%
%  \includegraphics[width=0.5\textwidth]{figures/Fig_forecast_flex_customer_1_training_from2016-04-01_to_2016-05-31.pdf}%
 % \label{}
%}%

\centering
 \begin{minipage}[t]{0.99\textwidth}
\subcaption{Summer}{%
  \includegraphics[width=0.99\textwidth]{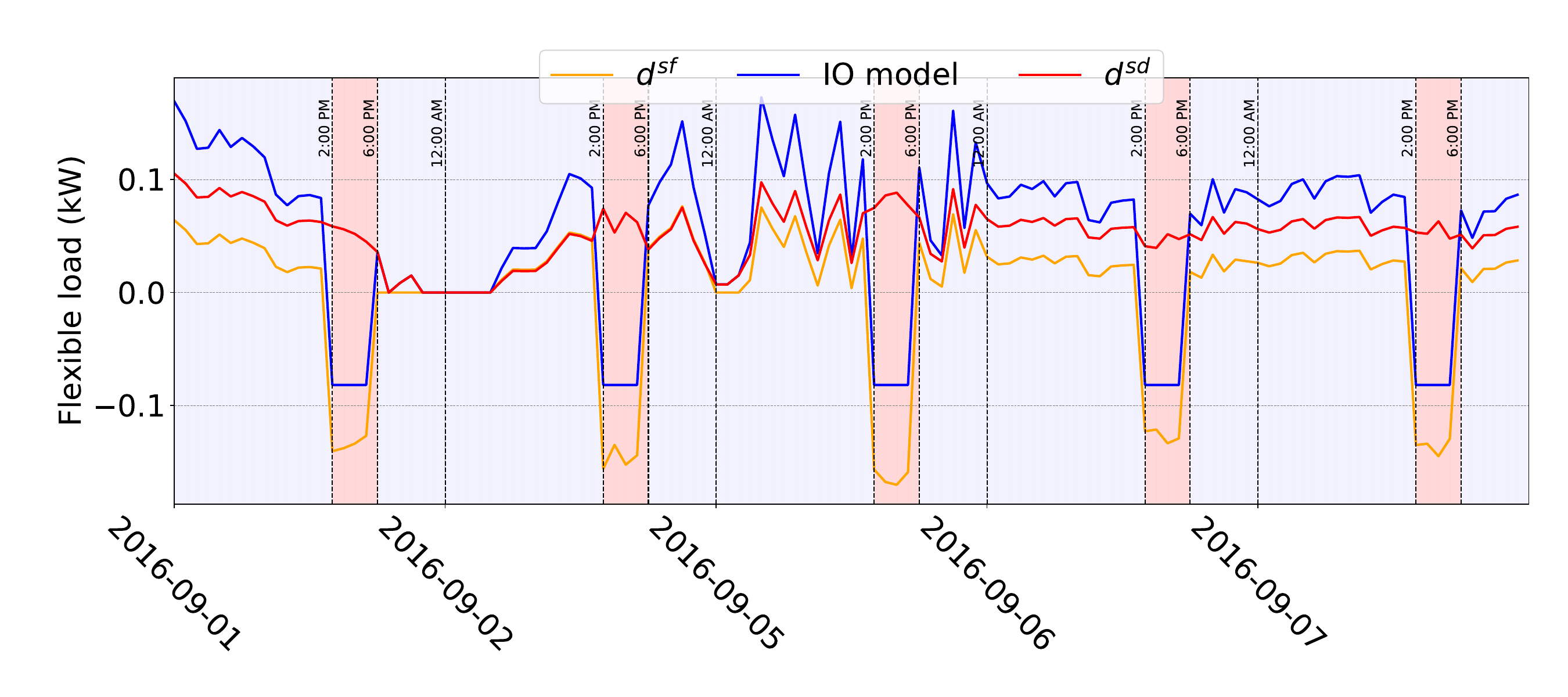}%
  \label{}
}%
\end{minipage}
%\subfloat[Season 4]{%
%  \includegraphics[width=0.5\textwidth]{figures/Fig_forecast_flex_customer_1_training_from2016-10-03_to_2016-11-30.pdf}%
%  \label{}
%}%

\caption{Flexible load profiles and performance error metrics  under best hyperparameter tuning}\label{Results_case_flex_load}
\end{figure}

\begin{table}[t]
\caption{Error metrics for Winter}
\label{tab:SEASON1}
\begin{tabular*}{\textwidth}{@{\extracolsep{\fill}}lcccccc}
\toprule
& \multicolumn{2}{c}{IO} & \multicolumn{2}{c}{XGboost} & \multicolumn{2}{c}{SARIMAX} \\
\cmidrule{2-3} \cmidrule{4-5} \cmidrule{6-7}
& MAE & RMSE & MAE & RMSE & MAE & RMSE \\
\midrule
Net load & 0.2077 & 0.2654 & 0.3737 & 0.4500 & 0.3720 & 0.4561 \\
$d^{bl}$ & - & - & 0.1270 & 0.1776 & 0.0659 & 0.0841 \\
$d^{flex}$ & - & - & 0.0711 & 0.0973 & 0.0589 & 0.0844 \\
Net load 2 & 0.2077 & 0.2654 & 0.2080 & 0.2592 & 0.2127 & 0.2727 \\
\hline
\end{tabular*}
\end{table}

\begin{table}[t]
\caption{Error metrics for Summer}
\label{tab:SEASON3}
\begin{tabular*}{\textwidth}{@{\extracolsep{\fill}}lcccccc}
\toprule
& \multicolumn{2}{c}{IO} & \multicolumn{2}{c}{XGboost} & \multicolumn{2}{c}{SARIMAX} \\
\cmidrule{2-3} \cmidrule{4-5} \cmidrule{6-7}
& MAE & RMSE & MAE & RMSE & MAE & RMSE \\
\midrule
Net load & 0.1778 & 0.2326 & 0.2889 & 0.3472 & 0.2507 & 0.2859 \\
$d^{bl}$ & - & - & 0.2682 & 0.2170 & 0.1406 & 0.1632 \\
$d^{flex}$ & - & - & 0.0728 & 0.0983 & 0.0579 & 0.0842 \\
Net load 2 & 0.1778 & 0.2326 & 0.2783 & 0.3221 & 0.2087 & 0.2396 \\
\hline
\end{tabular*}

\end{table}

%\begin{table}[t]
%\caption{Error metrics for Season 4}
%\label{tab:SEASON4}
%\centering
%\setlength\tabcolsep{3pt} % default value: 6pt
%{\color{black}\begin{tabular}{ccccccccccccc}
%\hline
%\multirow{3}{*}{} &  & \multicolumn{2}{c}{IO} & & \multicolumn{2}{c}{XGboost} & & \multicolumn{2}{c}{SARIMA} & & \multicolumn{2}{c}{SARIMAX} \\
%\cline{3-4} \cline{6-7} \cline{9-10} \cline{12-13}
% &    & MAE & RMSE  && MAE & RMSE  && MAE & RMSE && MAE & RMSE \\
%\hline
%    &  Net load  & 0.1835 &  0.2717 & & 0.2758 & 0.3506 & & 0.2014 &  0.2726 & & 0.2185 &  0.2935 \\
%    &  $d^{bl}$ & - &  - & & 0.0051 & 0.0090 & & 0.0344&  0.0608 & & 0.0725 &  0.0900 \\
%        &  $d^{flex}$  & - &  - & & 0.1200 & 0.1358 & & 0.0596 &  0.0944& & 0.0510 &  0.0874 \\
%    &  Net load 2 & 0.1835 &  0.2717 & & 0.1916 & 0.2807 & & 0.1676 &  0.2464 & & 0.1903 &  0.2646 \\
%\hline
%\end{tabular}}
%\end{table}

Finally, whilst the error metrics for the point forecasts validate the relative accuracy of the IO approach for forecasting, it is becoming increasingly important in demand forecasting to understand the tail risks in predictions. This involves producing density forecasts and assessing the calibrations of the predicted quantiles. For this purpose, we report the conventional  continuous rank probability scores (CRPS) averaged over the forecast horizon $T^{fo}$. 

\begin{table}[t]
\caption{Error metrics for the net load probabilistic forecasting: Quantiles and CRPS}
\label{tab:CRPS}
\begin{tabular*}{\textwidth}{@{\extracolsep{\fill}}lcccccccccc}
\toprule
& \multicolumn{3}{c}{IO} & \multicolumn{3}{c}{XGboost} & \multicolumn{3}{c}{SARIMAX} \\
\cmidrule{2-4} \cmidrule{5-7} \cmidrule{8-10}
& $q_{0.05}$ & $q_{0.95}$ & CRPS & $q_{0.05}$ & $q_{0.95}$ & CRPS & $q_{0.05}$ & $q_{0.95}$ & CRPS \\
\midrule
Winter & 0.171 & 1.136 & 0.232 & 0.494 & 1.334 & 0.329 & 0.783 & 0.8015 & 0.387 \\
Summer & 0.165 & 0.883 & 0.141 & 0.291 & 1.106 & 0.193 & 0.587 & 0.634 & 0.258 \\
\hline
\end{tabular*}
\end{table}
Table~\ref{tab:CRPS} contains the  $0.05,0.95-$quantiles, $q_{0.05}, q_{0.95}$, and the CRPS (averaged over the forecast horizon).  Note that lower scores are better and we observe that the IO method outperforms very substantially across all seasons and SARIMAX generally have intermediate performance.
Overall, the superior fitting and forecasting accuracy of the IO method for Net load appears to be well established against conventional benchmarks, in terms of both average accuracy and predictive density calibration. This established the feasibility and overall merit of the approach. The next question (Q4) is whether the estimated components really do represent what they are intended to represent, i.e., their construct validity.

To assess the construct validity of the flexible and baseload components delivered by the IO method, we regress each estimated component time series on the known metered appliance activities to assess their relative impact on the component's variation. We use the conventional  SHAP  (SHapley Additive exPlanations, \cite{shapley17}) values to represent the relative feature importance of the appliance activities.  SHAP is a technique for analyzing the explainability of machine learning models by providing a way to understand how a specific prediction is arrived at. In this way, we can see if the IO method provides meaningful load decompositions. Figures \ref{shapley_baseload} and \ref{shapley_flexible} display the relative feature strength of each appliance on the  baseload and flexible  components.  The features are ranked by their importance.  
Figure \ref{shapley_baseload} reveals that the baseload component primarily depends on habitual consumption patterns rather than discretionary appliance usage. This is evidenced by the dominance of autoregressive lags across both seasons, especially lag 1, which exhibits a strong positive relationship with baseload consumption. However, a key seasonal difference emerges: while lag 1 consistently increases baseload consumption in both seasons, the impact of other lagged terms is reversed (lags  6, 4 and
2 in winter and lags 4, 2 and 6 in summer), revealing that higher values contribute to lower baseload consumption.
Additionally, during the summer season, the "Living room" feature plays a distinct role, indicating that higher consumption levels in this area lead to increased baseload demand. This suggests a potential correlation with sustained cooling or ventilation use. 
For the flexible load components, a contrasting pattern is observed. {\color{black}Figures \ref{shapley_flexible}  and \ref{shapley_sheddable} reveal clear seasonal differences in the drivers of flexible consumption components.
For the shiftable load, Figure \ref{shapley_flexible} indicates a predominance of discretionary appliances, particularly the Dishwasher and the Furnace, whose SHAP values exceed those of most autoregressive terms. This suggests that winter shifting behavior is mainly triggered by appliance-level decisions rather than persistence effects. In contrast, during summer, the first autoregressive term (lag 1) becomes the most influential feature, surpassing individual appliances. This indicates a stronger temporal dependence in summer shifting dynamics.

For the sheddable load (Figure \ref{shapley_sheddable}), autoregressive effects are even more pronounced. In winter, lag 1 already appears among the dominant drivers, alongside the Furnace and Dishwasher. In summer, the dominance of autoregressive terms (lagds 1,2 and 3) becomes even clearer, while appliance-specific variables contribute comparatively less. This suggests that sheddable consumption is structurally more persistent and less directly associated with isolated appliance activation, especially in summer.

Additionally, the Barn feature shows noticeable importance during winter in the shiftable component, while its relevance declines in summer. This seasonal variation may reflect winter-specific usage patterns linked to barn-related activities. 
Overall, the SHAP analysis highlights a structural distinction: (i) Winter shiftable load is more appliance-driven, and (ii) Summer shiftable load and sheddable load exhibit stronger autoregressive behavior. This reinforces the difference between discretionary shifting behavior and more persistent, structurally constrained sheddable consumption.
These findings support the claim that the IO results effectively decompose total load into meaningful baseload and flexible components.
}

\begin{figure}
\centering
 \begin{minipage}[t]{0.49\textwidth}
\subcaption{Winter}{%
  \includegraphics[width=0.99\textwidth]{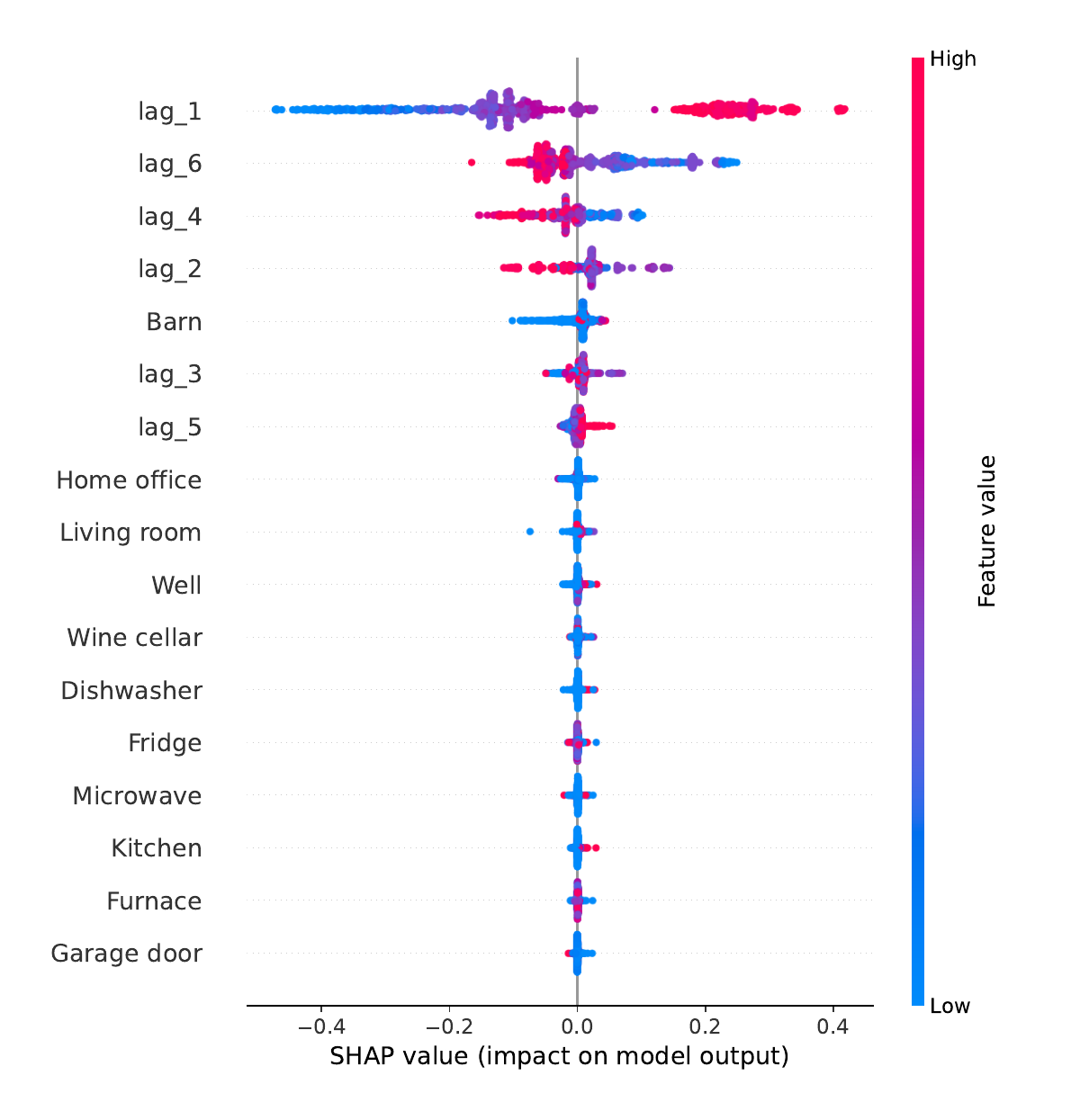}%
  \label{}
}%
\end{minipage}
%\subfloat[Season 2]{%
%  \includegraphics[width=0.5\textwidth]{figures/shapley_ie_1_training_from2016-04-01_to_2016-05-31.pdf}%
 % \label{}
%}%
\centering
 \begin{minipage}[t]{0.49\textwidth}
\subcaption{Summer}{%
  \includegraphics[width=0.99\textwidth]{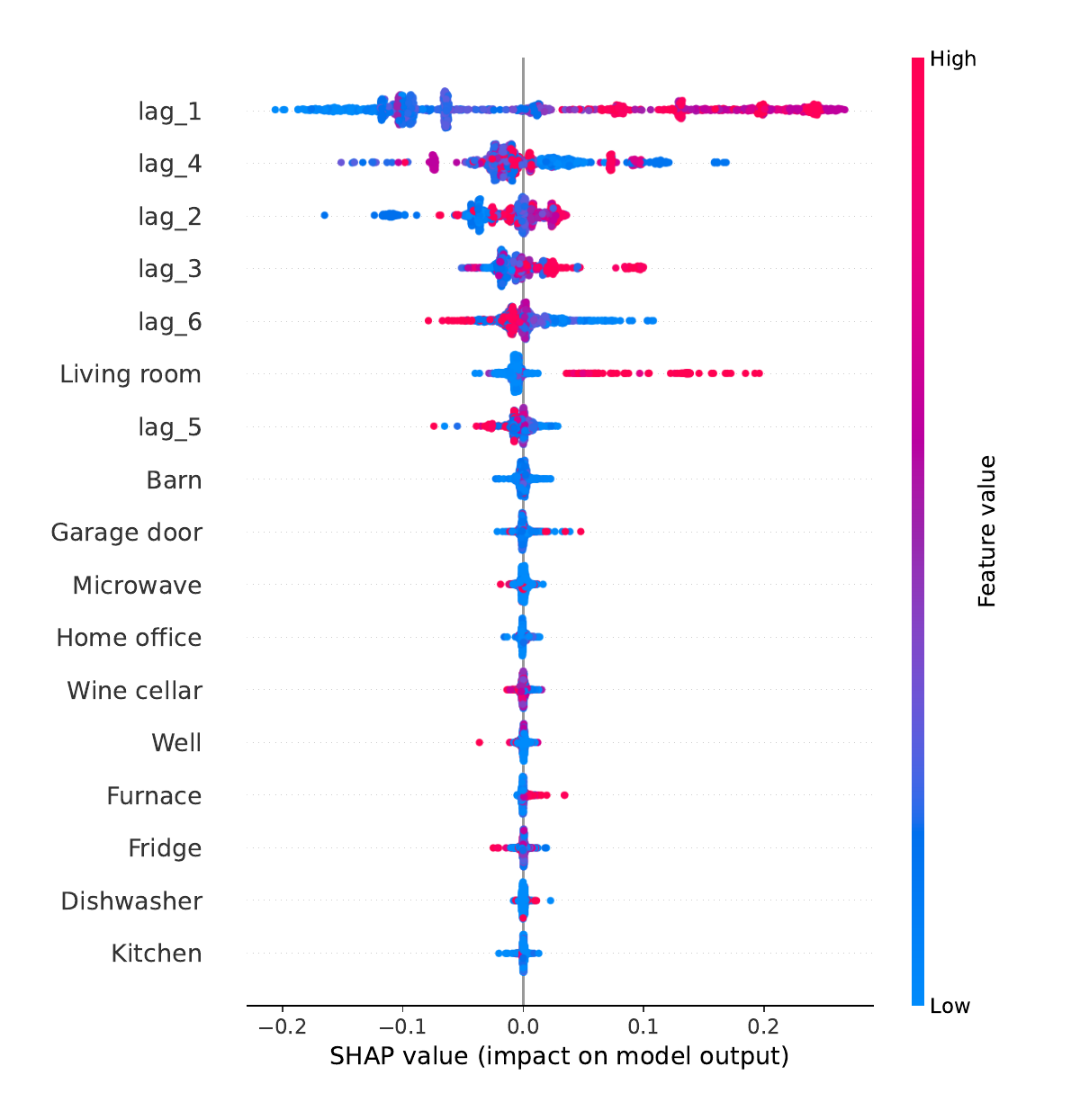}%
  \label{}
}%
\end{minipage}
%\subfloat[Season 4]{%
%  \includegraphics[width=0.5\textwidth]%{figures/shapley_ie_1_training_from2016-10-03_to_2016-11-30.pdf}%
%  \label{}
%}%

\caption{Time series' appliances impact into net baseload demand IO estimates via SHAP values}\label{shapley_baseload}
\end{figure}

\begin{figure}
\centering
 \begin{minipage}[t]{0.49\textwidth}
\subcaption{Winter}{%
  \includegraphics[width=0.99\textwidth]{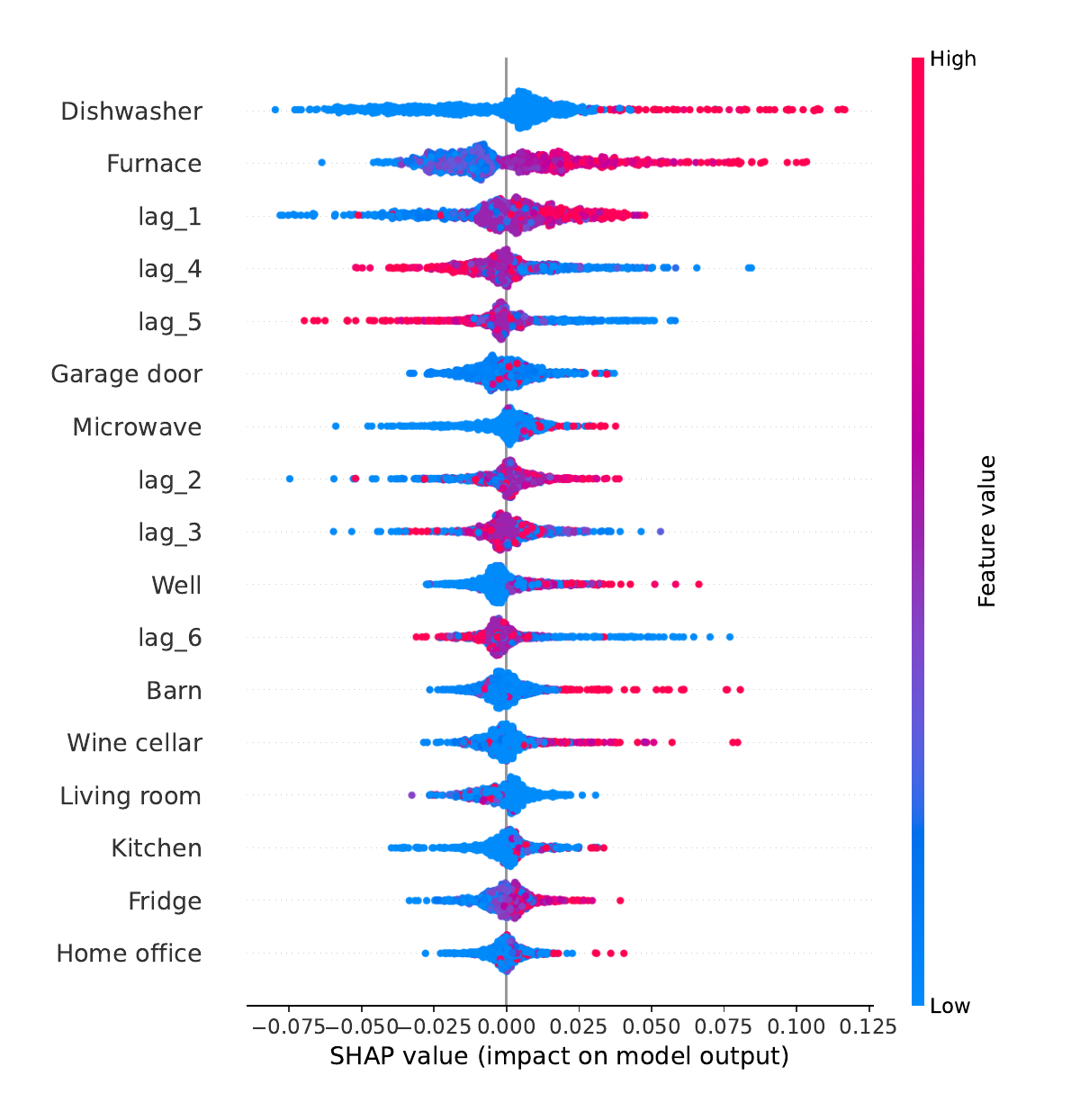}%
  \label{}
}%
\end{minipage}
%\subfloat[Season 2]{%
%  \includegraphics[width=0.5\textwidth]{figures/shapley_flex_1_training_from2016-04-01_to_2016-05-31.pdf}%
 % \label{}
%}%
\centering
 \begin{minipage}[t]{0.49\textwidth}
\subcaption{Summer}{%
  \includegraphics[width=0.99\textwidth]{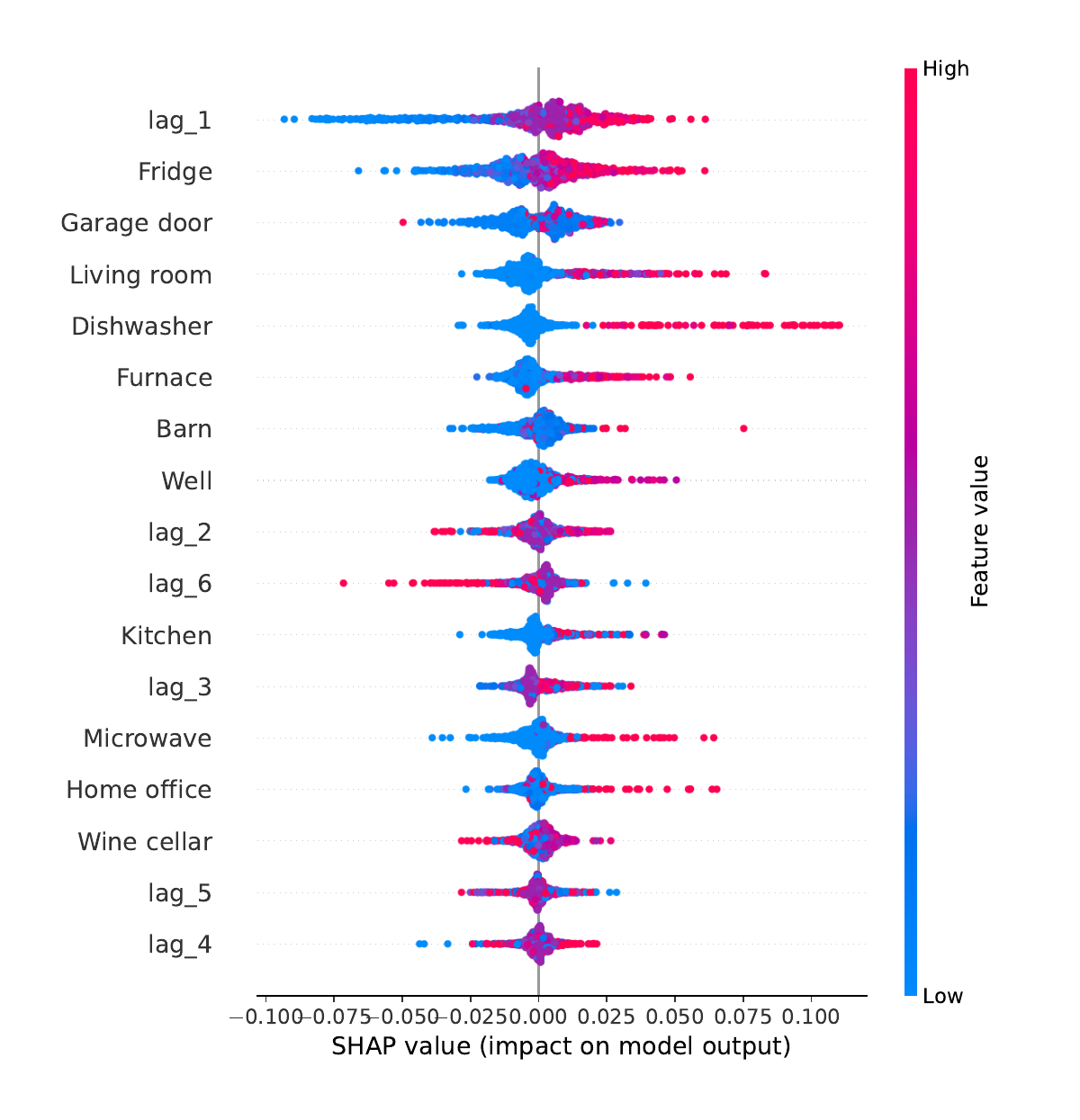}%
  \label{}
}%
\end{minipage}
%\subfloat[Season 4]{%
%  \includegraphics[width=0.5\textwidth]%{figures/shapley_flex_1_training_from2016-10-03_to_2016-11-30.pdf}%
%  \label{}
%}%

\caption{Time series' appliances impact into shiftable IO estimates via SHAP values}\label{shapley_flexible}
\end{figure}

\begin{figure}
\centering
 \begin{minipage}[t]{0.49\textwidth}
\subcaption{Winter}{%
  \includegraphics[width=0.99\textwidth]{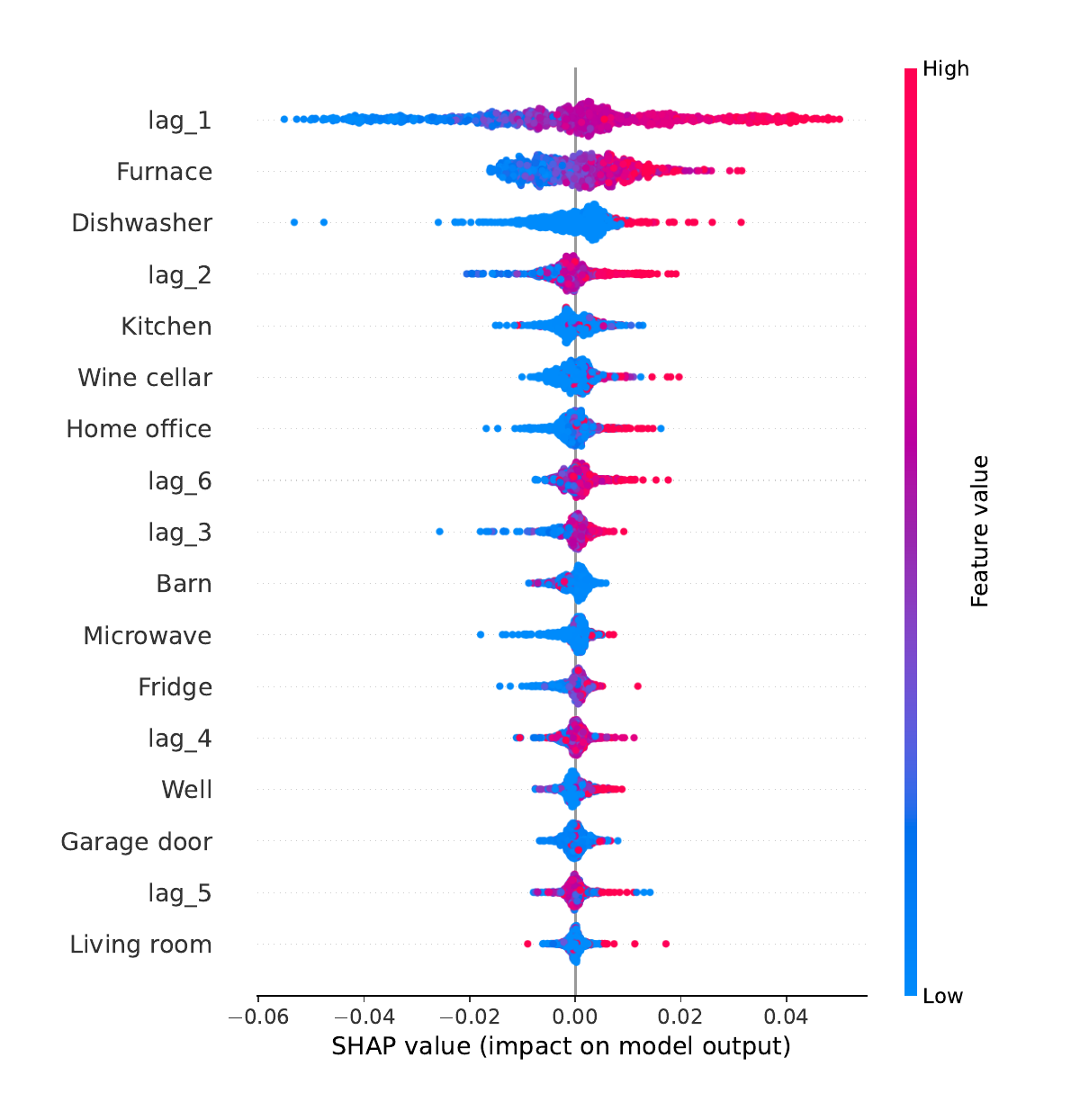}%
  \label{}
}%
\end{minipage}
%\subfloat[Season 2]{%
%  \includegraphics[width=0.5\textwidth]{figures/shapley_flex_1_training_from2016-04-01_to_2016-05-31.pdf}%
 % \label{}
%}%
\centering
 \begin{minipage}[t]{0.49\textwidth}
\subcaption{Summer}{%
  \includegraphics[width=0.99\textwidth]{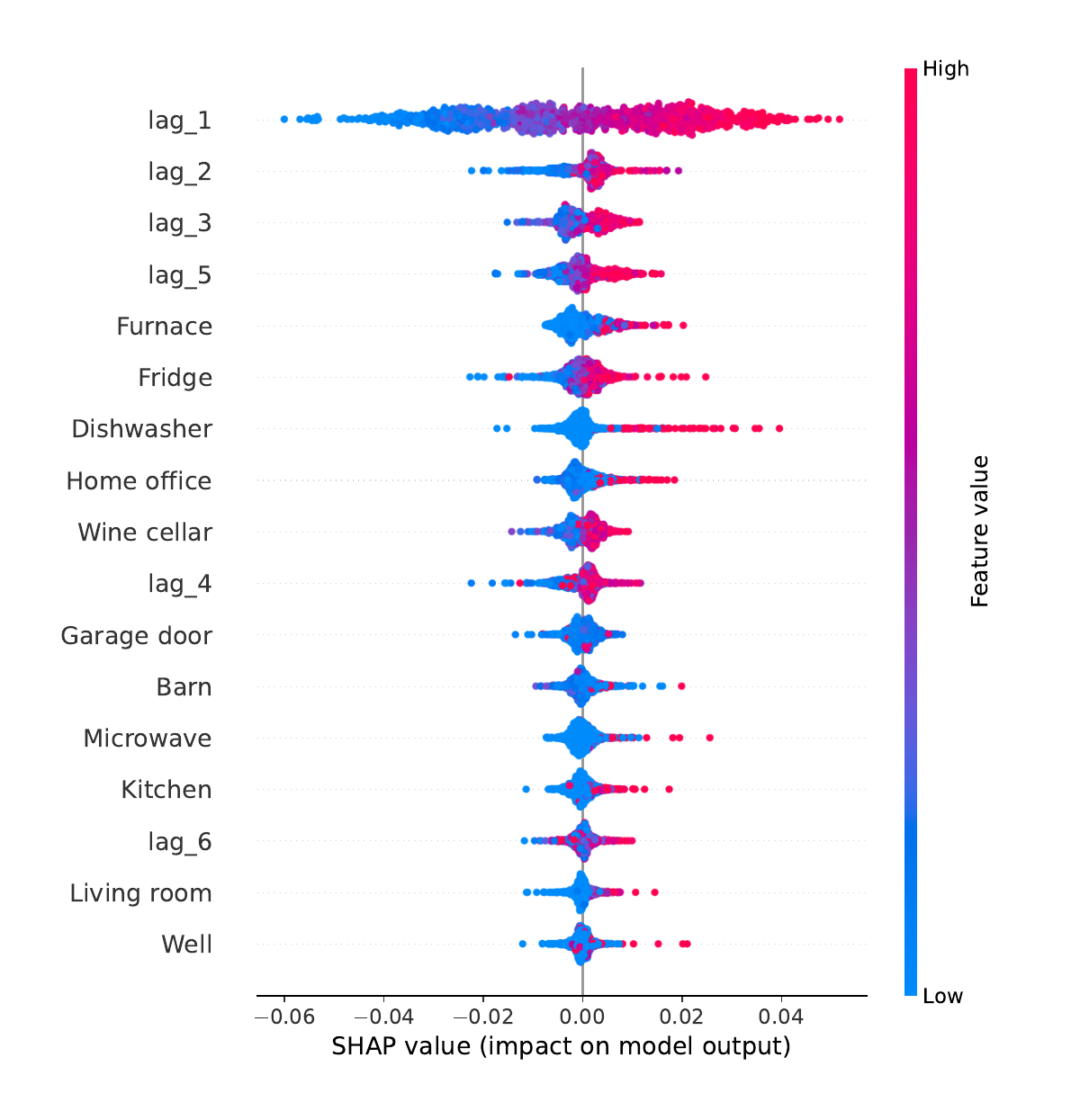}%
  \label{}
}%
\end{minipage}
%\subfloat[Season 4]{%
%  \includegraphics[width=0.5\textwidth]%{figures/shapley_flex_1_training_from2016-10-03_to_2016-11-30.pdf}%
%  \label{}
%}%

\caption{Time series' appliances impact into sheddable IO estimates via SHAP values}\label{shapley_sheddable}
\end{figure}

\section{Conclusions}\label{conclusions}

{\color{black}In this paper, we addressed an emerging information management challenge in retail electricity: how to characterize the latent components of domestic electricity demand response when only net demand at the meter is observable and behind-the-meter appliance-level information is unavailable or restricted. To overcome this limitation, we proposed a data-driven inverse optimization (IO) methodology that infers the unobservable flexible and baseload components of demand from aggregate net demand under structural behavioral assumptions, without requiring device-level measurements. Our experimental structure identified five key research questions, \textbf{RQ1-RQ5}, all of which were usefully addressed and summarised here, as follows:

(\textbf{RQ1}) Our results demonstrate that the proposed IO formulation is both computationally tractable and effective in recovering economically meaningful latent components from net demand alone. The estimated decomposition consistently produces distinct flexible and baseload profiles that align with the assumed behavioral structure of domestic demand. This establishes a proof of concept: the IO framework can reliably infer unobservable demand components using only aggregate observations and a structural optimization model.

(\textbf{RQ2}) We find that the latent components identified by the IO model are predictable out of sample using exogenous variables such as weather, calendar effects, and autoregressive terms. Importantly, this predictability is achieved without any knowledge of appliance activities. Benchmark comparisons with statistical and machine learning models show that the IO-derived components exhibit comparable or superior out-of-sample predictability, indicating that the decomposition captures stable and systematic drivers of demand behavior rather than noise.

(\textbf{RQ3}) Empirical forecasting results show that reconstructing net demand by forecasting and summing the IO-estimated flexible and baseload components yields more accurate forecasts than modeling the aggregate net demand series directly. This confirms the practical value of the decomposed approach: separating structurally different demand drivers improves forecast precision by capturing heterogeneous dynamics that are otherwise masked in the aggregate signal (RQ3).

(\textbf{RQ4}) Construct validity is supported by strong and interpretable statistical relationships between the IO-derived components and observed appliance activities. SHAP-based regressions reveal that discretionary appliance usage is significantly associated with the flexible component, while autoregressive effects and habitual consumption patterns dominate the baseload component. These findings indicate that the latent variables recovered by IO correspond to real underlying behavioral processes, rather than being purely statistical artifacts.

(\textbf{RQ5}) Overall, the IO-based decomposition leads to more accurate and behaviorally consistent estimates of price responsiveness compared to conventional aggregate benchmark models. By explicitly modeling flexible and baseload demand as latent optimization-driven components, the framework captures nonlinear and heterogeneous price responses that are obscured in aggregate approaches. This provides stronger empirical support for the proposition that IO offers a practically useful and interpretable tool for estimating consumer demand response in retail electricity markets.

Taken together, these findings show that inverse optimization provides a viable and policy-relevant solution to the lack of behind-the-meter data. The methodology enables accurate inference of latent demand structure, improves forecasting performance, and yields economically meaningful estimates of price response without requiring intrusive data collection. From a practical perspective, the resulting model-driven decision support system can enhance coordinated decision-making among key power system stakeholders. Grid operators can better anticipate local demand flexibility and manage constraints, while retailers gain improved tools for designing incentives that unlock consumer-side flexibility. By aligning operational planning with behavioral insights, the proposed IO framework positions demand flexibility as a measurable and actionable resource, supporting a more sustainable, reliable, and cost-efficient electricity system.

}
Beyond the main findings, several promising directions for future research arise. The framework may be further developed by incorporating richer behavioral models, capturing consumer heterogeneity, and utilizing real-time or IoT-based data where available. In addition, applying the IO methodology across a range of market structures, regulatory environments, and geographic contexts would help evaluate its generalizability and resilience. Lastly, future work could examine how the inferred demand components can be effectively embedded into pricing mechanisms, demand response programs, and grid management practices, thereby unlocking their full practical and economic potential.

% Appendix here
% Options are (1) APPENDIX (with or without general title) or
%             (2) APPENDICES (if it has more than one unrelated sections)
% Outcomment the appropriate case if necessary
%
% \begin{APPENDIX}{<Title of the Appendix>}
% \end{APPENDIX}
%
%   or
%
\appendix
\section{Notation}\label{notation_appendix}
All notations is presented in Table~\ref{tab-notation}.
\begin{table}[h!]
\centering
\footnotesize
\begin{tabular}{p{5em}p{40em}} 
\toprule
   \textbf{Notation} & \textbf{Abbreviations} \\
 \midrule
 SO & System operator\\
  TOU & Time-Of-Use\\
   IO & Inverse optimization\\
   FOP & Forward optimization program\\
   BTM & Behind-the-meter \\
   GDPR,SLP &  General Data Protection Regulation,  Successive Linear Programming\\
 \midrule 
 \textbf{Notation} & \textbf{Indices and Sets} \\
 \midrule
     $[T]$ & Review period set, defined as $[T]=\{1, \ldots, T\}$ made by $T$ hours and indexed by
    $t$\\
     $[S]$ & Set of $S$ days defined as $[S]=\{1, \ldots, S\}$ and indexed by
      $s$\\
    $\Theta$ & Flexibility feasibility set\\
  \midrule
 \textbf{Notation} & \textbf{Exogenous variables and parameters} \\
 \midrule
  $\hat{\mathbf{d}}_{s}$ & Vector $T$-dimensional of load measurements on day $s$ (kWh)\\
    $\hat{\mathbf{g}}_s$ & Vector of $T$-dimensional renewable energy generation on day $s$ (kWh)\\
        $p_t$ & Price/flat tariff at time $t$ (c/kWh, JPY/kWh)\\
 $p^{sf,+}_{s,t}$ & Per unit price incentive of energy of shifting  by increasing at time $t$ on day $s$  (c/kWh, JPY/kWh)\\
 $p^{sf,-}_{s,t}$ & Per unit price incentive of energy of shifting  by decreasing at time $t$ on day $s$ (c/kWh, JPY/kWh)\\
 $p^{sd}_{s,t}$ & Per unit price incentive of energy of shedding at time $t$ on day $s$ (c/kWh, JPY/kWh)\\
          $\mathbf{z}_{s}$ & Vector of $T$-dimensional exogenous observations build
up by  price on day $s$ \\
         $\hat{\boldsymbol{\xi}}_{s,t}$ & Vector of external features at time $t$ on day $s$ (kWh)\\     
                  $\alpha$, $\omega_s$ & Forgetting factor parameter,  Normalized forgetting  weight  on day $s$\\
   $c^{sd}_{s,t}$ & Comfort cost incurred by load shedding at time $t$ on day $s$ (c/$\text{kWh}^2$, JPY/$\text{kWh}^2$)\\
      $c^{sf,+}_{s,t}$ & Comfort cost incurred by load shifting by increasing demand at time $t$ on day $s$ (c/$\text{kWh}^2$, JPY/$\text{kWh}^2$)\\
            $c^{sf,-}_{s,t}$ & Comfort cost incurred by load shifting by decreasing demand at time $t$ on day $s$ (c/$\text{kWh}^2$, JPY/$\text{kWh}^2$)\\
                     $\mathbf{c}_{s}$ & Vector of $T$-dimensional exogenous observations build
up by cost on day $s$ \\
                 $M_0, \underline{M}$ & Bound for the intercept and maximum capacity, respectively (kWh)\\
                        $[\beta^{flex}_0, \boldsymbol{\beta}^{flex}]$ & Affine decision rule parameters\\
                $\ell$ & Utility function\\
 \midrule
 \textbf{Notation} & \textbf{Endogenous variables} \\
 \midrule
 $d^{bl}_{t}$ & Baseload demand at time $t$ (kWh)\\
$d^{sf,-}_{s,t}, d^{sf,+}_{s,t}$ & Downward/Upward shifted demand at time $t$ on day $s$ (kWh)\\
$d^{sf}_{s,t}$ & Shifted demand at time $t$ on day $s$ (kWh)\\
$d^{sd}_{s,t}$ & Sheddable demand part of the net demand at time $t$ on day $s$ (kWh)\\
$\overline{d}^{flex}_{s,t}$ & Flexibility envelope function at time $t$ on day $s$ (kWh)\\
%$\overline{d}^{sf,+}_{s,t},\overline{d}^{sf,-}_{s,t}$ & Shiftability envelope functions at time $t$ on day $s$ (kWh)\\
$d^{sd,-}_{s,t}$ & Sheddable demand at time $t$ on day $s$ (kWh)\\
$d^{flex}_{s,t}$ & Flexible demand at time $t$ on day $s$ (kWh)\\
%$\delta^{sf,+}_{s,t},\delta^{sf,-}_{s,t}$ & Auxiliary binary variables at time $t$ on day $s$ (kWh)\\
$\mathbf{u}_s$ &Array of demand attributes $(d^{bl}_{t}, \overline{d}^{sf,+}_{s,t}, \overline{d}^{sf,-}_{s,t},\overline{d}^{sd}_{s,t})$ on day $s$\\
        $\boldsymbol{\theta}_{s}$ & Array of decision variables of the $s$-th FOP on day $s$\\
  $\theta_{t}$ & Array of decision variables $(d^{sf,+}_t,d^{sf,-}_t,d^{sd,-}_t)$ of the consumer's problem at time $t$\\
\bottomrule
\end{tabular}
\caption{\bf Notation}
\label{tab-notation}
\end{table}
\section{Computational details}
Computations were performed on a MacBook Pro equipped with an Apple M4 chip (10-core CPU) and 24 GB of Unified Memory using the solver MOSEK under Pyomo \citep{pyomo}.
The IO model depends on two hyperparameters, namely: $param=[ \alpha,\lambda ]$,  which were tuned via a grid-search. For the SARIMAX, we choose the parameters that minimize the AIC pmdarima package via autoarima function. XGboost is applied by using the skforecast Python package \citep{skforecast} via default options with 6 lags.
\subsection{Performance metrics: MAE, RMSE and CRPS}\label{performance_metrics}
\begin{align}
    \text{MAE}(d^{true}, d^{fo})&=\dfrac{1}{|T^{fo}|}
    \displaystyle\sum_{t \in T^{fo}}|d^{true}_t-d^{fo}_t|\\
            \text{RMSE}(d^{true}, d^{fo})&=\sqrt{\dfrac{1}{|T^{fo}|}
    \displaystyle\sum_{t \in T^{fo}}(d^{true}_t-d^{fo}_t)^2}
\end{align}
where $d^{true}, d^{fo}$ are the load measurements and forecast values, and  $T^{fo}$ is the planning horizon set (in our case, 5 days of 24 hours). We assume that on every day at hour $t$  the external regressors
are perfectly known.  This is often assumed in forecasting research in order to focus more directly upon the comparative methodologies so that the results are not confounded by estimation errors in variables.

 CRPS is a strictly proper scoring rule which is  widely used within density  forecasting research.  It  needs to be approximated in practice, and one approach is based on exploiting its equivalent quantile-based representation \cite{Gneiting2011}:
\begin{equation}\label{CRPS_formula}
  \text{CRPS}(F,y)=2\int_0^1 \text{QL}_q(F^{-1}(q),y)\text{d}q  
\end{equation}
where $F,F^{-1}$ are the distribution and quantile function, respectively, $y$ is the measurement, and QL is the quantile loss function.   Computing the proxy for CRPS  requires the $q$-quantiles, where $q\in \{0.01, 0.05, \ldots, 0.95, 0.99  \}$. We used the forecasting method with the pinball loss function (also known as quantile score function) available in \citep{skforecast}.
\subsection{Consumer peak and off-peak pricing in the case study}\label{data_IO_ohio_tokyo}
 We estimated a typical flat tariff of  $22$ c/kWh and prices of $29$ c/kWh and $15$ c/kWh for the peak and off-peak periods, with the peak defined as  $2-6$ pm in summer and $6-9$ am in winter.  We also define 
$p^{sf,+}_t=\max(p_t-\text{TOU}_t,0), p^{sf,-}_t=\max(\text{TOU}_t-p_t,0)$. The parameters for load shedding are defined by $\sum_{t\in [T]} p^{sf,+}_t / 24$ and the cost $c^{sd}$ is defined by  $5(p^{sd})^2$. The parameters $c^{sf,+}_t,c^{sf,-}_t$ are set to $|p_t-\text{TOU}_t|$ when $p^{sf,+}_t, p^{sf,-}_t$ are equal to zero, respectively, and $10^{-5}$, otherwise. {\color{black}We take as (standardized) regressors the temperature difference, $\text{Tdiff}:=T-T^{app}$, the solar generation, and the two Cartesian wind velocity components, corresponding to the zonal (east–west) and meridional (north–south) projections of wind speed derived from the wind bearing angle, for the flexible envelope decision rule}.

%\item[{\bf Tokyo Case Study.}] We estimated a typical flat tariff of  $22$ c/kWh and prices of $29$ c/kWh and $15$ c/kWh for the peak and off-peak periods, with the peak defined as  $2-6$ pm in summer and $6-9$ am in winter.  We also define 
%$p^{sf,+}_t=\max(p_t-\text{TOU}_t,0), p^{sf,-}_t=\max(\text{TOU}_t-p_t,0)$. The parameters for load shedding are defined by $\sum_{t\in [T]} p^{sf,+}_t / 24$ and the cost $c^{sd}$ is defined by  $p^{sd}$. The parameters $c^{sf,+}_t,c^{sf,-}_t$ are set to $|p_t-\text{TOU}_t|$ when $p^{sf,+}_t, p^{sf,-}_t$ are equal to zero, respectively, and $10^{-5}$, otherwise. For $p_t$ in the IO model the flat tariff was $26\, \text{JPY/kWh}$ and we defined
%$p^{sf,+}_t=\max(p_t-\text{TOU}_t,0), p^{sf,-}_t=\max(\text{TOU}_t-p_t,0)$, where $\text{TOU}_t=35\, \text{JPY/kWh}$ between 2pm and 10pm,  and $20 \,\text{JPY/kWh}$ otherwise.
%The price incentive for load shedding, $p^{sd}$, is defined by $\sum_{t\in [T]} p^{sf,+}_t / 24$ and the cost $c^{sd}$ is given by  the price incentive $p^{sd}$. The parameters $c^{sf,+}_t,c^{sf,-}_t$ are set to $|p_t-\text{TOU}_t|$ when $p^{sf,+}_t, p^{sf,-}_t$ are equal to zero, respectively, and 0, otherwise. 

\section{Single-level conic reformulation Supplement }\label{sec:reformulation}
The Karush-Kuhn-Tucker (KKT) conditions for the lower-level problem enables us to reformulate the (regularized) IO  model as a single-level optimization program:

\begin{subequations}  \label{IO:reform_conic}
\begin{align}
\min &  \displaystyle \sum_{s \in [S]}\omega_s u_s+\lambda  t&  \label{obj:conic_reform}\\
 \text{ \text{s.t.} }   
 & u_s \geqslant
\left\|\left( 
\mathbf{d}^{bl}+\mathbf{d}^{sf}_{s} + \mathbf{d}^{sd}_{s}-\hat{\mathbf{g}}_{s}-\mathbf{\widehat{d}}_{s}\right)_s \right\|^2_2  & \forall s \in [S]\\
& t \geqslant\left\|\boldsymbol{\beta}^{flex} \right\|^2_2,\; t \geqslant 0\\
  & \eqref{eq:elastic}-\eqref{eq:compact_feas_region}\\
&  -2c^{sf,+}_{s,t}d^{sf,+}_{s,t}+p^{sf,+}_{s,t}-p_{s,t}+\kappa_{s}-\mu^{+}_{s,t}+\nu^+_{s,t}=0&  \forall   s \in [S] ,\forall t \in [T] \\
&-2c^{sf,-}_{s,t}d^{sf,-}_{s,t}+ p^{sf,-}_{s,t}+p_{s,t}-\kappa_{s}-\mu^{-}_{s,t}+\nu^-_{s,t}=0&  \forall   s \in [S], \forall t \in [T] \\
&-2c^{sd}_{s,t}d^{sd,-}_{s,t}+  p^{sd}_{s,t}+p_{s,t}-\mu^{0}_{s,t}+\nu^{0}_{s,t}=0&  \forall   s \in [S] ,\forall t \in [T] \\
&  \mu^+_{s,t} (\beta^{flex}_{0}+
    \langle \boldsymbol{\beta}^{flex} , \hat{\boldsymbol{\xi}}_{s,t} \rangle-d^{sf,+}_{s,t})=0&  \forall   s \in [S], \forall t \in [T] \\
&  \mu^{-}_{s,t} (\beta^{flex}_{0}+
    \langle \boldsymbol{\beta}^{flex} , \hat{\boldsymbol{\xi}}_{s,t} \rangle-d^{sf,-}_{s,t})=0&  \forall   s \in [S], \forall t \in [T] \\
&  \mu^{0}_{s,t} (\beta^{flex}_{0}+
    \langle \boldsymbol{\beta}^{flex} , \hat{\boldsymbol{\xi}}_{s,t} \rangle-d^{sd,-}_{s,t})=0&  \forall   s \in [S], \forall t \in [T] \\
&  \nu^{0}_{s,t} d^{sd,-}_{s,t}=0&  \forall   s \in [S],\forall t \in [T] \\
&  \nu^+_{s,t}d^{sf,+}_{s,t}=0&  \forall   s \in [S] ,\forall t \in [T] \\
&  \nu^{-}_{s,t}d^{sf,-}_{s,t}=0&  \forall   s \in [S], \forall t \in [T] \\
&  \mu^{+}_{s,t}, \mu^{-}_{s,t}, \mu^{0}_{s,t},  \nu^+_{s,t}, \nu^{-}_{s,t} , \nu^{0}_{s,t}\geqslant 0&  \forall   s \in [S], \forall t \in [T] \\
&  \boldsymbol{\theta}_s \in \Theta(\beta^{flex}_{0}+
    \langle \boldsymbol{\beta}^{flex} , \hat{\boldsymbol{\xi}}_{s,t} \rangle)   &  \forall s\in [S]   \\
& \kappa_{s} \in \mathbb{R}, u_s \geqslant 0&  \forall   s \in [S]  \label{last_eq:conic_reform}
\end{align}
\end{subequations}
 The above program can be recast as a  mixed-integer program with linear objective function and $S+1$ rotated conic quadratic constraints  solvable by off-the-shelf conic optimization software such as MOSEK \citep{mosek} version 11.1.6.

\bibliographystyle{elsarticle-num-names}

\bibliography{References} % if more than one, comma separated

% CASE 2: BiBTeX used to generate mypaper.bbl (to be further fine tuned)
%\input{mypaper.bbl} % outcomment this line in Case 2

%%%%%%%%%%%%%%%%%
\end{document}